\documentclass[structabstract]{aa}  
\usepackage{amsmath}
\usepackage{natbib}
\usepackage{units}
\usepackage{color}

\usepackage{graphicx}
\usepackage{txfonts}

\newcommand{\beq}{\begin{equation}}
\newcommand{\eeq}{\end{equation}}

\begin{document}
	\title{Modelling Circumbinary Planets: The case of Kepler-38}
	\author{Wilhelm Kley
 		\inst{1}
	\and
		Nader Haghighipour
 		\inst{1,2}
	}

	\institute{
	Institut f\"ur Astronomie und Astrophysik, Universit\"at T\"ubingen, Auf der Morgenstelle 10, D-72076 T\"ubingen, Germany. \\
	\email{wilhelm.kley@uni-tuebingen.de}
	\and
        Institute for Astronomy and NASA Astrobiology Institute, University of Hawaii-Manoa, Honolulu, HI 96825, USA. \\
	\email{nader@ifa.hawaii.edu}
	}

 	\date{Received ; accepted }

\abstract
        {Recently, a number of planets orbiting binary stars have been discovered by the {\textit {Kepler}} space telescope.
    In a few systems the planets reside close to the dynamical stability limit.
    Due to the difficulty of forming planets in such close orbits, it is believed that they have formed further out in the disk
    and migrated to their present locations.}
        {Our goal is to construct more realistic models of planet migration in circumbinary disks, and to determine 
         the final position of these planets more accurately. In our work, we focus on the system Kepler-38 where 
         the planet is close to the stability limit.
 	}
 	{The evolution of the circumbinary disk is studied using two-dimensional hydrodynamical simulations. 
        We study locally isothermal disks as well as more realistic models that include full viscous heating, radiative cooling from the 
        disk surfaces, and radiative diffusion in the disk midplane.
        After the disk has been brought into a quasi-equilibrium
        state, a 115 Earth-mass planet is embedded and its evolution is followed.
        }
	{In all cases the planets stop inward migration near the inner edge of the disk.
         In isothermal disks with a typical disk scale height of $H/r = 0.05$, the final outcome agrees
         very well with the observed location of planet Kepler-38b. 
         For the radiative models, the disk thickness and location of the inner edge is
         determined by the mass in the system. For surface densities in the order of 3000 g/cm$^2$ at 1 AU, the inner gap lies close to
         the binary and planets stop in the region between the 5:1 and 4:1 mean-motion resonances with the binary. 
         A model with a disk with approximately a quarter of the mass yields a final position very close to the observed one.
        }
        {For planets migrating in circumbinary disks, the final position is dictated by the structure of the disk.
         Knowing the observed orbits of circumbinary planets, radiative disk simulations with embedded planets can provide
         important information on the physical state of the system during the final stages of its evolution.
        }
   \maketitle
%

\section{Introduction}

An interesting development in exoplanetary science has been the recent discovery of circumbinary planets (CBPs) by the 
{\it Kepler} space telescope. In contrast to systems where a planet revolves around one star of a binary,
here the planet orbits the entire binary system. Currently known main sequence binaries with circumbinary planets are: 
Kepler-16 \citep{2011Sci...333.1602D}, Kepler-34 and 35 \citep{2012Natur.481..475W}, 
Kepler-38 \citep{2012ApJ...758...87O}, Kepler-47 \citep{2012Sci...337.1511O} 
and Kepler-64 \citep{2013ApJ...768..127S}. In all these systems, the binary is close with an orbital period
of 7 to 41 days. The orbital periods of their planets range from 50 to 300 days.

Since the discovery of the first circumbinary planet in the Kepler 16 system, it has been noted that
in some of these systems, the innermost planet is very close to the boundary of the dynamical 
stability \citep{1986A&A...167..379D,1999AJ....117..621H}. Given that, as indicated by the observations,
the orbital planes of these binaries are perfectly aligned with those of their planets, this implies that
the planets are formed in flat, circumbinary disks. The proximity of the orbits of these CBPs to the 
stability limit then raises the question as to whether these planets formed in their current orbits, or
at farther distances from the binary and migrated to their present locations.

Although not all aspects of planet formation are fully understood, it is widely accepted that planets 
form around single stars
through a bottom-up process where growth is achieved via a sequence of sticking collisions with subsequent gas accretion 
for the more massive objects \citep{2010apf..book.....A}. In a binary star system,
the presence of the binary in the middle of the protoplanetary disk will alter this process and make planet formation more 
complicated. The perturbation of the binary and its (eccentric) disk will dynamically excite the orbits of planetesimals 
and hinder their growth to larger objects \citep{2007MNRAS.380.1119S,2012ApJ...761L...7M, 2013A&A...553A..71M}. 
Because in a circumbinary disk, the outer edge of the central cavity, that is created due to the effect of tidal 
forces from the binary on the disk material, coincides with the boundary of the planetary stability, the previous statement
implies that the in situ formation of CBPs close to the stability limit may not be possible \citep{2012ApJ...754L..16P}. 
However, because of the small separations of CBP-hosting binaries (each of the above-mentioned systems fits into 
the orbit of Mercury), their effects on the formation of planets at large distances will be negligibly small
suggesting that these planets could have
formed farther out and migrated to their current orbits \citep{2013MNRAS.435.2328D, 2013A&A...553A..71M}.


Planet migration
is a natural 
consequence of planet-disk interaction \citep{2000MNRAS.318...18N, 2012ARA&A..50..211K}. 
To study the migration of planets in circumbinary disks, the structure of the disk has to be analyzed
and compared to those around single stars. The most important dynamical difference between the former and latter disks
is the existence of a central cavity in the circumbinary disks. 
As shown by \citet{1994ApJ...421..651A}, the radial extent of this cavity is a function of
the binary semi-major axis, eccentricity, and mass-ratio as well as the disk viscosity. As indicated by these authors, 
for typical values of the disk viscosity, and depending on the binary eccentricity, the location of the inner edge of the 
disk will be about 2 to 3 times the separation of the binary. 

The migration of planets in circumbinary disks was first studied by \citet{2003MNRAS.345..233N} for massive, Jupiter-type planets.
He found that for small values of the binary eccentricity, the migrating planet will be captured in a 
4:1 mean-motion resonance (MMR) with the binary
whereas in more eccentric systems ($e_\text{bin} \gtrsim 0.2$), the planet would be captured in a stable orbit further out
\citep[see also][]{2008A&A...483..633P}. Subsequent studies by \citet{2007A&A...472..993P} extended these analyses to planets
with masses as small as 20 $M_\text{Earth}$ showing that these planets usually stop near the edge of the cavity,
and they suggested that planets in circumbinary disks should be predominantly found in that 
area. As the inner edge of the disk roughly coincides with the boundary of planetary stability,
the predictions of these authors turned out to be in a very good agreement with the orbital architecture of several Kepler CBPs.
More general cases in which accretion and multiple planets were also considered were later studied by the same authors
\citep{2008A&A...478..939P,2008A&A...483..633P}.

In a recent paper \citet{2013A&A...556A.134P} revisited planet migration in circumbinary disks and developed models to
explain the orbits of the planets around binaries Kepler-16, Kepler-34 and Kepler-35. Similar to the majority of the 
simulations of this kind, these authors considered a locally isothermal approximation where the disk thermodynamics 
is modeled by prescribing a given temperature profile. They also considered a closed boundary condition at the edge 
of the cavity assuming zero mass accretion onto the central binary. Although the work by \citet{2013A&A...556A.134P}
represents significant results, their applicability may be limited due to the simplifying assumptions of an isothermal 
disk with a closed boundary conditions for the central cavity. Numerical simulations by 
\citet{1996ApJ...467L..77A} and \citet{2002A&A...387..550G} have shown that despite the appearance of the above-mentioned cavity in the 
center of a circumbinary disk, material can still flow inside and onto the central binary. In other words, a more realistic
boundary condition would allow for the in-flow of the material through the inner edge of the disk into the disk cavity.


Here, we present an improved and extended disk model that allows for much more realistic simulations. 
We consider the system of Kepler-38 as this system represents one of the binaries in which the circumbinary
planet is in close proximity to the stability limit. 
We have developed an improved approach which includes detailed balance of viscous heating and radiative cooling from the surface
of the disk \citep{2003ApJ...599..548D}, as well as additional radiative diffusion in the plane of the disk \citep{2008A&A...487L...9K}.
For a planet embedded in disks, this improved thermodynamics can have dramatic effects on the planet orbital dynamics such that
for low-mass planets, it can even reverse the direction of migration \citep{2008A&A...487L...9K,2009A&A...506..971K}. 
Also, unlike \citet{2013A&A...556A.134P}, we allow free in-flow of material from the disk into the central cavity and construct 
models with net mass in-flow through the disk.


This paper is organized as follows. In Section 2, we present the physical setup of our disk models. In Section 3,
we study locally isothermal disks and in Section 4, we present the results of our full radiative models and compare 
them to the results from standard isothermal cases. Section 5 concludes
this study by summarizing the results and discussing their implications.

\section{The hydrodynamic model}
As mentioned above, we consider Kepler-38 to be our test binary star and carry out simulations for that system.
To model the evolution of a disk around a binary star, we consider a circumbinary disk around Kepler-38, and
assume that the disk is vertically thin and the system is co-planar. The assumption of co-planarity is well
justified as the mutual inclination of Kepler-38 and its planet is smaller than $0.2$ degrees. 
We then perform two-dimensional (2D) hydrodynamical simulations in the plane of the binary.

To simulate the evolution of the disk, the viscous hydrodynamic equations are solved in a polar coordinate system 
($r,\phi$) with its origin at the center of mass of the binary. In our standard model, which we use as a basis 
for comparing the results of our subsequent simulations, we consider the radial extent of the disk $(r$) to be 
from 0.25 to 2.0 AU, and $\phi$ to vary in an entire annulus of $[0,360^\circ]$.
This domain is covered by an equidistant grid of $256 \times 512$ gridcells. For testing purposes, we have also
used a grid twice as fine. In all models we evolve the vertically integrated equations for the surface density $\Sigma$, 
and the velocity components ($v_r, v_\phi$). 

When simulating locally isothermal disks, we do not evolve the energy equation, and instead use an isothermal 
equation of state for the pressure. When considering radiative models,
a vertically averaged energy equation is used, which evolves the temperature of midplane 
\citep{2012A&A...539A..18M}. Radiative effects are include in two ways. 
First, a cooling term is considered to account for the radiative loss from the disk surface
\citep{2003ApJ...599..548D}. Second, we include diffusive radiative transport in the midplane of the disk 
using flux-limited diffusion \citep{2008A&A...487L...9K}. The use of a flux-limiter is required
because near its inner edge, the disk is very optically thin, which would lead to an unphysical 
high energy flux in the diffusive part of radiative transport equations. For more details on the implementation of
this diffusive transport, see \citet{2013A&A...560A..40M}.
In our radiative simulations, we consider the full, vertically averaged dissipation \citep{2012A&A...539A..18M}.
However, stellar irradiation is not taken into account.

To calculate the necessary height of the disk $H$ at a position $\mathbf r$, we first note that
in a circumstellar disk (around a single star at the position ${\bf r}_\ast$), the vertical height of the disk is given by
\beq
\label{eq:diskheight-single-star}
H({\bf r}) =c_{\rm s}\left({\frac{\left|{\bf r}-{\bf r_\ast}\right|^{3}}{G M_\ast}}\right)^{-\frac{1}{2}}\,.
\eeq
In this equation, $M_\ast$ is the mass of the central star,  $c_s$ is the speed of sound in the disk at the location $\vec{r}$,
and $G$ is the gravitational constant. In a binary star system, $H(\mathbf r)$ has to be calculated by taking   
the contributions of both stars into account \citep{2002A&A...387..550G}. That is,
\beq
  \label{eq:diskheight}
  H({\bf r}) =
  \left(\sum_{i=1,2}{\frac{G M_i}
     {c_{\rm s}^2\sqrt{|{\mathbf r} - {\mathbf r_i}|}^3}}
  \right)^{-\frac{1}{2}}
   =  \left(\sum_{i=1,2}{H_i^{-2} (\vec{r})}
  \right)^{-\frac{1}{2}} \,,
\eeq
where ${M_i}={M_1}$ and $M_2$ are the masses of the stars of the binary.
Equation~\ref{eq:diskheight} indicates that in a circumbinary disk, the total height $H$ is always smaller than the individual heights $H_i$.

The equation of state of the gas in the disk is given by the ideal gas law using a mean molecular weight, $\mu =2.35$ (in atomic mass units),
and an adiabatic exponent of $\gamma = 1.4$.
For the shear viscosity we use the $\alpha$-parametrization with $\alpha = 0.01$ for our standard model, and we set the bulk viscosity
to zero. For the Rosseland opacity, we use analytic formula as provided by \citet{1985prpl.conf..981L},
and the flux-limiter as in \citet{1989A&A...208...98K}. 

To calculate the gravitational potential of the binary-planet system at a position $\vec{r}$ in the disk, 
we use, for all 3 bodies, a smoothed potential function of the form
\begin{equation}
  \Psi_\text{k} (\vec{r})  = - \frac{G M_k}{ [ (\vec{r} - \vec{r}_k)^2 + (\epsilon H)^2]^{1/2}} \,, 
\end{equation}
where $M_k$ denotes the masses of the objects (stars and planet),
$\vec{r}-\vec{r}_k$ is the vector from a point in the disk to each object, and 
$H$ is calculated using eq.~(\ref{eq:diskheight}).
The quantity $\epsilon$ in eq. (3) is a smoothing parameter that accounts for the effect of the finite thickness of the disk. 
In our simulations, we assume  $\epsilon = 0.6$ or $0.7$ \citep{2012A&A...541A.123M}. We note that
for the planet, the smoothing length cannot be smaller than $\epsilon \, r_\text{H}$, 
where $r_\text{H}=a_\text{p}(m_\text{p}/3M_\text{bin})^{1/3}$ is the radius of the Hill sphere around the planet,
$a_\text{p}$ is the semi-major axis of the planet’s orbit, and $M_\text{bin} = M_1 + M_2$.
The indirect terms are taken into account through a shift of the positions of the objects such that the
binary's barycenter always coincides with the origin of the coordinate system.
 
The hydrodynamical equations are integrated using a mixed, explicit/implicit, scheme. The standard hydrodynamical
terms are integrated explicitly using a standard Courant condition with Courant number $f_\text{C} = 0.66$.
The viscous and the diffusive radiative parts are integrated implicitly to avoid instabilities.
For the time-integration of the hydrodynamical equations, we utilize the {\tt RH2D}-code with the FARGO 
\citep{2000A&AS..141..165M} upgrade.

The motions of the stellar binary and the planet are integrated using a 4th order Runge-Kutta integrator.
All objects, including the two stars, feel the gravity of the disk as well as their mutual gravitation. 
The self-gravity of the disk is not considered. 

To calculate the force from the disk acting on the planet, we exclude parts of the Hill sphere of the planet
using a tapering function
\beq
f(d)=\left[\exp\left(-\frac{d/r_\text{H}-p}{p/10}\right)+1\right]^{-1}\,.
\label{eq:tapering}
\eeq
In this equation, $d$ is the distance from the grid cell to the planet, and
$p$ is a dimensionless parameter that is set to $0.6$ for the isothermal models and $0.8$ for the radiative runs. 
We calculate the orbital parameters of the planet  
using Jacobian coordinates, assuming the planet orbits a star with mass $M_\text{bin} = M_1 + M_2$, at the binary
barycenter.

\subsection{Initial setup and boundary conditions}
\label{subsec:setup}
\begin{table}[t]
\caption{The binary parameter and the observed planetary parameter of the Kepler-38 system
 used in the simulations.
 The mass of the primary star is $0.949 M_\odot$. The values have been
 taken from \citet{2012ApJ...758...87O}.
 \label{tab:kepler38}
}
a) Binary Parameter \\
\medskip
\begin{tabular}{l|l|l|l} 
\hline
 Mass ratio   &  Period  &  $a_\text{bin}$   &   $e_\text{bin}$   \\  
   $q = M_2/M_1$  &  [days]   &    [AU]       &           \\
\hline
  0.2626          &   18.6  &  0.1469    & 0.1032    \\
\hline
\end{tabular} 

\medskip

b) Planet Parameter \\
\medskip
\begin{tabular}{l|l|l|l}
\hline
 Mass   &  Period  &  $a_\text{p}$   &   $e_\text{p}$   \\
   $M_\text{Jup}$  &  [days]   &    [AU]       &           \\
\hline
  0.34          &   105.6  &  0.46    &  $<$ 0.03    \\
\hline
\end{tabular}

\end{table}

In all models, the disk initial surface density is chosen to have a $\Sigma (r) = \Sigma_0 \, r^{-1/2}$ profile where $r$ is the
distance from the center of mass of the binary, and $\Sigma_0 = 3000$g/cm$^2$ is the surface density at 1 AU. 
Such a radial profile for $\Sigma$ corresponds to that usually adopted for the minimum solar mass nebula.
As the total stellar mass in the Kepler-38 system is approximately 1.2 solar-masses, one can assume that, to a first 
order of approximation, the mass of the circumbinary disk would lie in the same range as that of the disk around the protosun.
Assuming that the observed circumbinary planets have reached their current positions in the late phase of
their evolution, such a disk surface density may be on the high side. However, together with the relatively large value
of the viscosity parameter ($\alpha = 0.01$), it allows for a sufficiently fast evolution of the system in our standard model
such that it can be numerically simulated. We will vary these parameters for the radiative models below.

We choose the initial temperature of the disk to vary with $r$ as $T(r) \propto r^{-1}$ such that, 
assuming a central star of $M_\text{bin}$, the vertical thickness of the disk will always maintain the condition $H/r = 0.05$.
The initial  angular velocity of the disk at a distance $r$ is chosen to be equal to the Keplerian velocity at that
distance, and the radial velocity is set to zero.

The parameters of the central binary have been adopted from \citet{2012ApJ...758...87O} and can be found in Table~\ref{tab:kepler38},
together with the planetary data.
Due to the interaction of the binary with the disk (and planet), these parameters will slowly vary during a simulation.
For the models with an embedded planet, we consider the planet to have a mass of $m_\text{p} = 3.63 \times 10^{-4} M_1$
which is equivalent to $m_\text{p} = 0.34 M_\text{Jup}$ or about 115 Earth-masses (see Table~\ref{tab:kepler38}).
At the beginning of each simulation, we start the binary at its periastron. 
 
The boundary conditions of the simulations are constructed such that at the outer boundary of the disk, $r_{max}=2.0$ AU,
the surface density remains constant.
This is achieved by using a damping boundary condition where the density is relaxed toward
its initial value, $\Sigma(r_{max}) = 2121 $g/cm$^2$, and the radial velocity is damped toward zero. For this, we use the procedure specified by
\citet{2006MNRAS.370..529D}. The angular velocity at the outer boundary is also kept at the initial Keplerian value,
and for the temperature we use a reflecting condition so that there will be no artificial radiative flux through the 
outer boundary. These conditions at the outer boundary lead to a disk with zero eccentricity at $r_{max}$. Hence,
$r_{max}$ has to chosen large enough such that the inner regions are not influenced.


At the inner boundary $(r_{min})$, we consider a boundary condition such that the in-flow of disk material onto the binary is allowed.
This means, for the radial boundary grid cells at $r_{min}$, we choose a zero-gradient mass out-flow condition, 
where the material can freely leave the grid and flow onto the binary. No mass in-flow into a grid is allowed at $r_{min}$.
The zero-gradient condition is also applied to the angular velocity of the material since due to the effect of the binary,
no well-defined Keplerian velocity can be found which could be used otherwise. This zero-gradient condition for the
angular velocity implies a physically more realistic {\it zero-torque} boundary.

With these boundary conditions, the disk can reach a quasi-stationary state in which there will be a
constant mass-flow through the disk.

\begin{table}
\caption{Properties for the locally isothermal disk models without a planet.
Model 1 is our reference model that is described in detail in Sect.~\ref{subsec:setup}.
In the other two models certain (numerical) parameter variations that have been applied in order to compare to
previous models, in particular by \citet{2013A&A...556A.134P}. They are described in the quoted sections.
 \label{tab:models}
}
\begin{tabular}{l|l|l}
 Name    &  Characterization & Section  \\             
\hline
 Model 1  &   standard model  &  in \ref{subsec:setup} \\
 Model 2  &   closed inner boundary  & \ref{subsec:closed-rmin} \\
 Model 3  &   different inner radii  &  \ref{subsec:rminc}  \\
\hline
\end{tabular}  
\end{table}

\section{The locally isothermal case}
\label{sec:isothermal}
In this section we do not evolve the energy equation but instead leave the temperature
of the disk at its initial value. This procedure has the advantage of making the simulations much faster
as no heating and cooling of the disk has to be considered.
This is an approximation that has often been used in planet disk simulations and gives the
first order results, but it also has its limitations when planetary migration is concerned \citep{2008A&A...487L...9K}.
We will first show the results of our standard case (model 1), and then 
compare these results to those recently obtained by \citet{2013A&A...556A.134P} who also used a locally isothermal model.
Table~\ref{tab:models} gives a very brief overview of the models and can serve as a reference.

We begin by discussing models without a planet where we bring the disk into equilibrium and
analyse its structure.
This equilibration process is required because the disk structure is determined by the action of
the binary for which no simple analytic model can be prescribed. Additionally, the evolution of the
planet in the disk occurs so slowly that the disk always remains near equilibrium.

\subsection{The structure of the circumbinary disk}
\label{subsec:iso-noplanet}
\begin{figure}
\center
\includegraphics[width=0.45\textwidth]{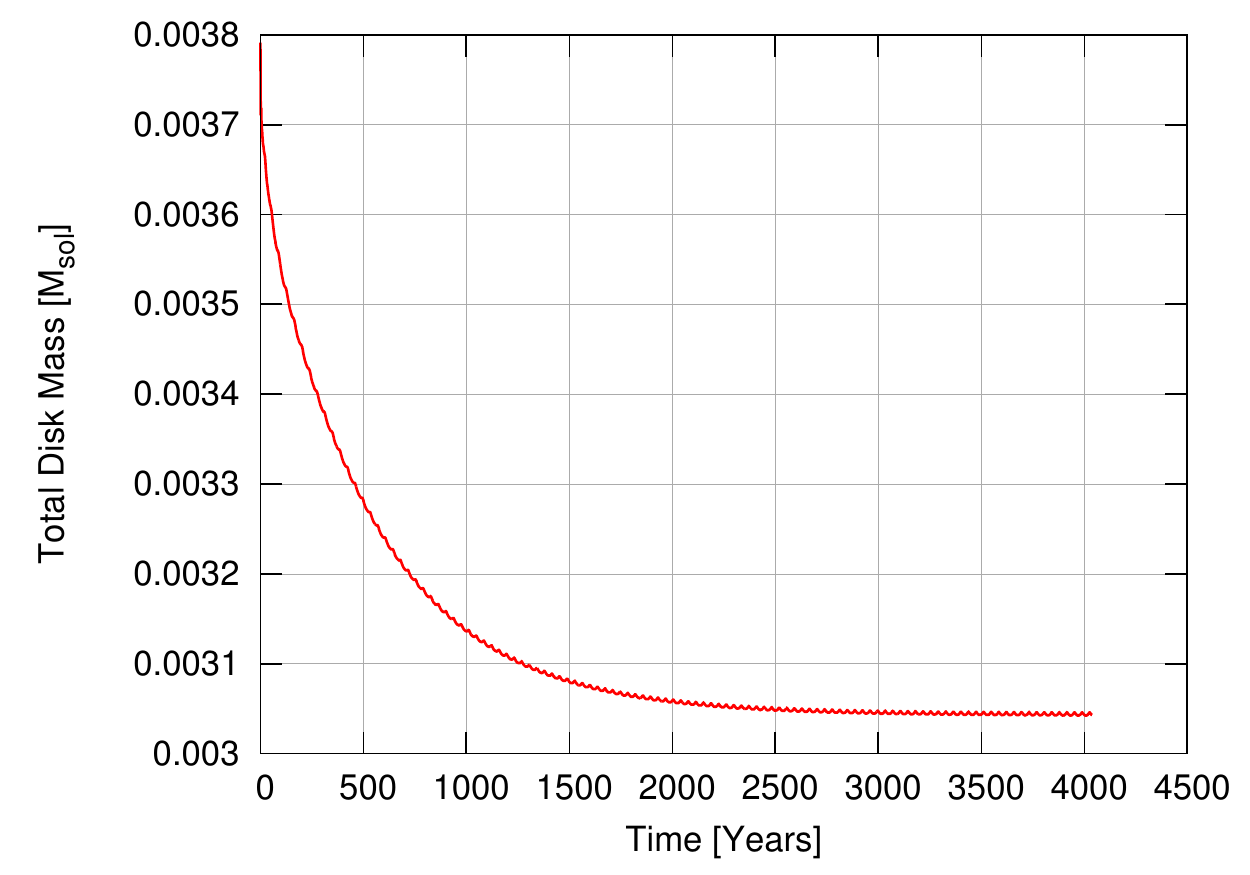} 
\caption{Graph of the total mass of the disk (without the planet) in our locally isothermal standard model (model 1). 
After a few thousand years, the disk reaches equilibrium and a nearly constant mass is established.
}
\label{fig:k38a-diskmass}
\end{figure}
\begin{figure}
\center
\includegraphics[width=0.45\textwidth]{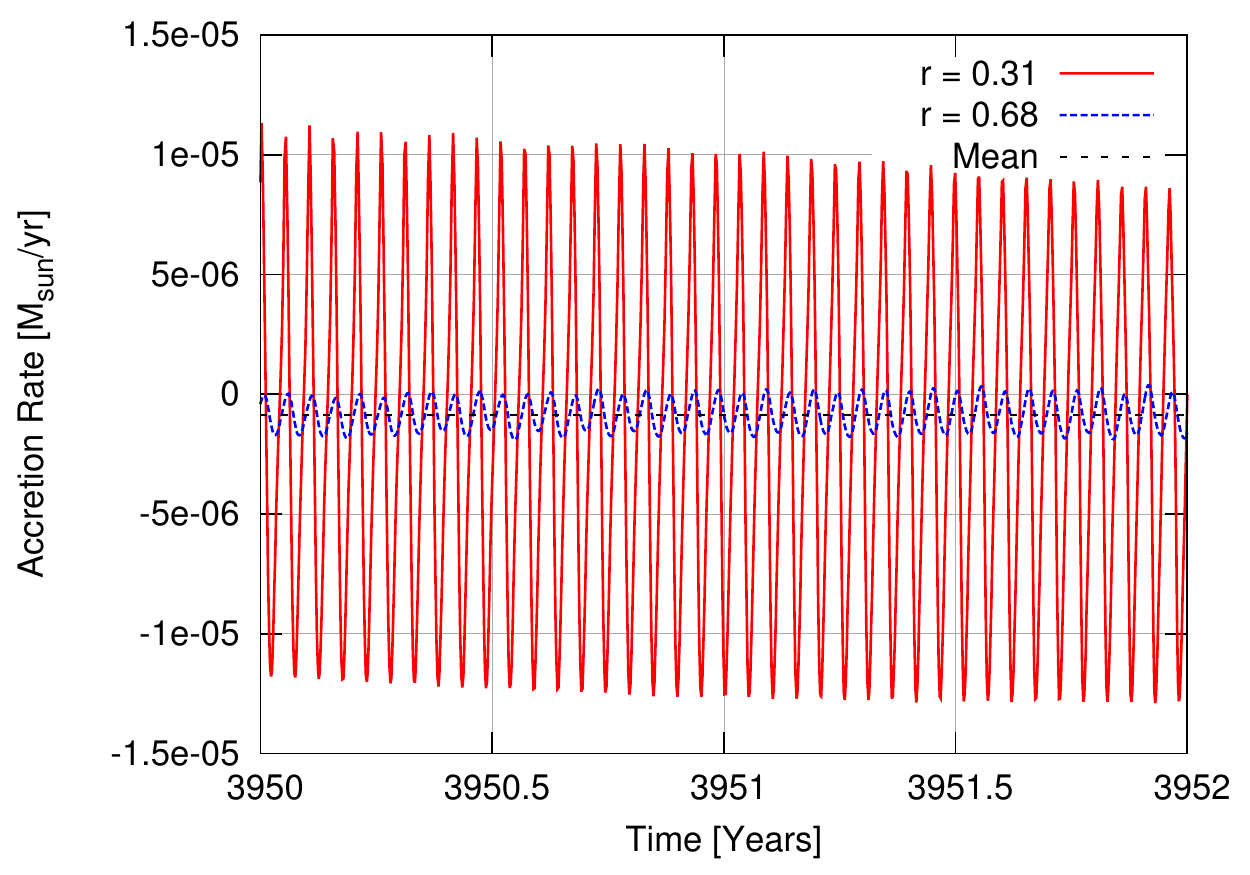}
\caption{Graph of the rate of mass accretion through the disk in model 1 measured at two different distances.
The mean value, $-8.68 \times 10^{-7}$, is also plotted for $r=0.68$ AU. 
As the central binary accretes material, the mass flow is directed inward and becomes negative. However,
in the main text, we report the accretion rate in form of positive absolute values.
}
\label{fig:k38a-mdot}
\end{figure}

\begin{figure}
\center
\includegraphics[width=0.45\textwidth]{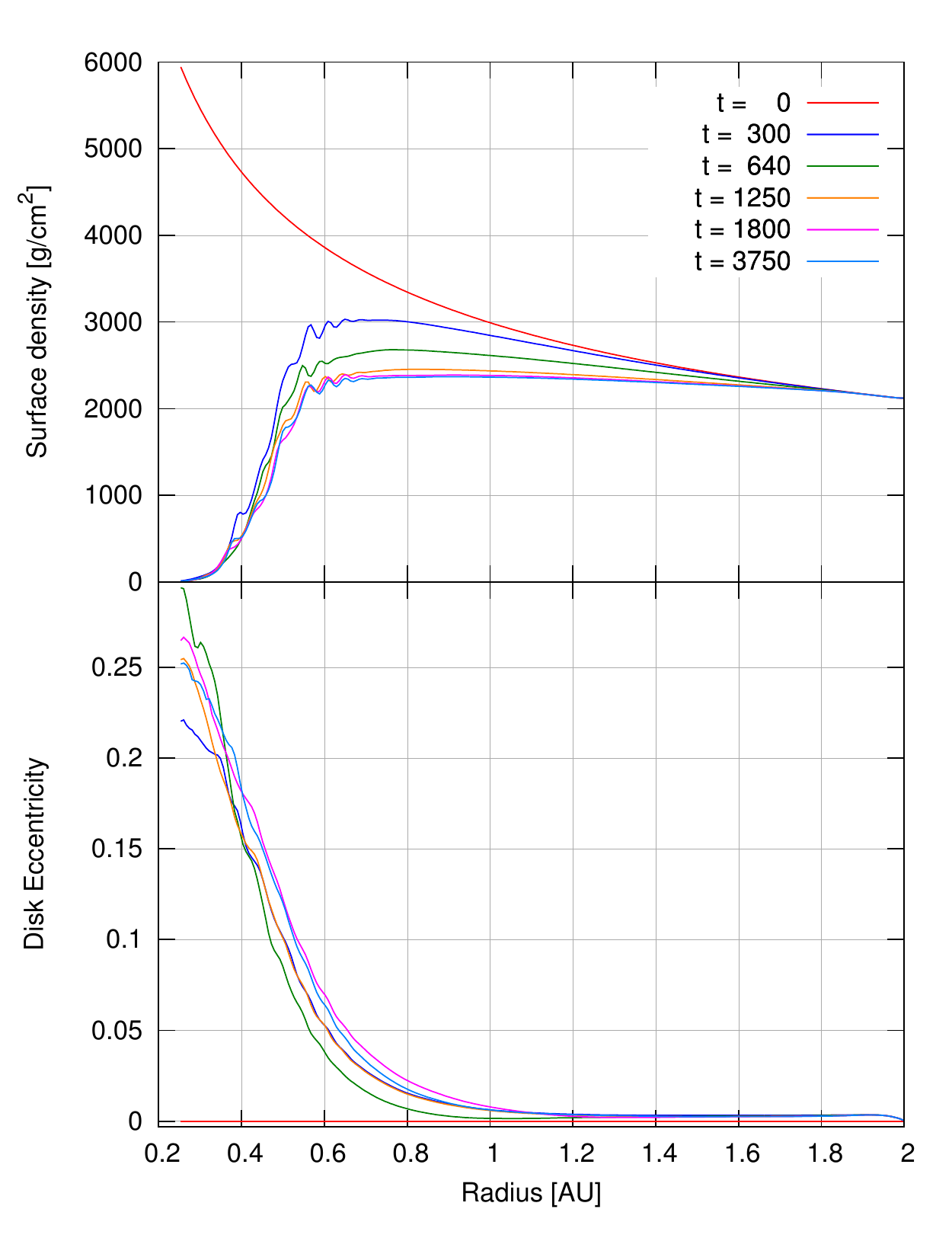} \\
\caption{Azimuthally averaged surface density and disk eccentricity for model 1.
Displayed are the profiles at different evolutionary times (in yrs), where the red curve denotes the initial
setup.
}
\label{fig:k38a-sigecc}
\end{figure}

To analyse the disk's own dynamics (i.e., {\it without} a planet) due to the effect of the central binary,
we simulated the dynamics of the disk in our standard model with a zero-mass planet.
Due to the fact that the disk inner boundary is open, 
and at the outer boundary, its surface density is considered to be constant, the disk settles into a final quasi-stationary 
state in which the mass of the disk is constant and there is a constant mass accretion rate onto the central binary.
This has been shown in  Fig.~\ref{fig:k38a-diskmass} where the total mass in the computational domain is plotted vs. time
for the locally isothermal model with no planet. After about 2000-3000 yrs a quasi-stationary state has been reached 
at which state the outflow through the inner boundary is exactly balanced by the 'inflow' from outside as established
by keeping the value of the density constant at $r_{max}$. During this process, the disk mass has decreased by about
$20$\% from $3.8 \times 10^{-3} M_\odot$ to $3.04 \times 10^{-3} M_\odot$. The small remaining oscillations are caused by the
eccentric motion of the inner binary which induces variations of the disk structure. In this simulation where we have been interested in the
structure of the disk without the planet, the disk does not affect the motion of the binary. Nevertheless, the motion of the
binary has been numerically integrated to examine the accuracy of the numerical method. We find that during the total evolution time
of about 4000 years, which corresponds to over 80,000 binary orbits and 4.5 million(!) time steps, the semi-major axis of the binary has
only shrunk by $0.014 \%$, a value sufficiently small for our purposes.

At equilibrium, the rate of the mass-flow through the disk, $\dot{M}_\text{disk}$, becomes on average nearly constant.  
Fig.~\ref{fig:k38a-mdot} shows this where $\dot{M}_\text{disk}$, measured at two different distances, is plotted near the
end of the simulation.
The resulting accretion rate equals approximately $\dot{M}_\text{disk} = 8.7 \times 10^{-7} M_\odot$/yr.
The oscillations are much larger in the inner parts of the disk due to the perturbation from the binary,
and they occur at the binary's orbital period.
This value may be compared to the typical theoretical equilibrium disk accretion rate,
$\dot{M}_\text{th} = 3 \pi \Sigma \nu$. At the outer boundary, where the mass density
has been kept constant, we find $\dot{M}_\text{th} = 5.4 \times  10^{-7} M_\odot$/yr.

During time, the surface density slowly evolves away from its initial profile
until it settles in a new equilibrium state. 
This is shown in the upper panel of Fig.~\ref{fig:k38a-sigecc} where the azimuthally
averaged radial surface density profile is shown at different evolutionary times. 
In agreement with Fig.~\ref{fig:k38a-diskmass}, the equilibration time takes about $2000$ yrs. 
At this state, the surface density profiles will no longer change with time. Because of the tidal action of the binary 
on the disk, a central gap is formed with a surface density many orders of magnitude smaller than inside the disk. 
The inner edge of the disk, which we can define, very approximately, as that radius at which
the surface density is about half the maximum value, lies here at around $r \approx 0.45$AU. This is slightly larger than
$3 a_\text{bin}$ and in a good agreement with the findings of \citet{1994ApJ...421..651A}.
We note that for the binary system considered here, the stability limit for planetary orbits is about 0.4~AU 
\citep{1986A&A...167..379D,1999AJ....117..621H}. Outside of the gap beyond $r=0.6$AU, the surface density 
profile becomes relatively flat. One can notice that the mass-flow rates shown in Fig.~\ref{fig:k38a-mdot}  
correspond to a region deep inside the gap ($r=0.31 \approx 2 a_\text{bin}$), and immediately outside of it ($r=0.68$).

As was shown by \citet{2013A&A...556A.134P}, circumbinary disks can attain significant eccentricities.
We calculate the eccentricity of the disk by treating each grid cell as an individual particle with
a mass and velocity equal to the mass and velocity of the cell \citep{2006A&A...447..369K}. 
To calculate a radial dependence for the disk eccentricity, $e_\text{disk}(r)$, we average over the angular direction.
In our simulations, the disk eccentricity remains small, as shown in the lower panel of Fig.~\ref{fig:k38a-sigecc}.
In the regions outside of the central gap (beyond $r=0.6$), the disk eccentricity is always 
below $e_\text{disk} < 0.07$. At radial distances $r > 1.0$, this eccentricity becomes smaller than about $0.01$. 
These values of the disk eccentricity are in contrast to those reported by \citet{2013A&A...556A.134P}, who found, 
on average, larger values.
We attribute this difference to the assumption of an open inner boundary,
and we will analyse this in more detail in the next section.

\begin{figure}
\center
\includegraphics[width=0.45\textwidth]{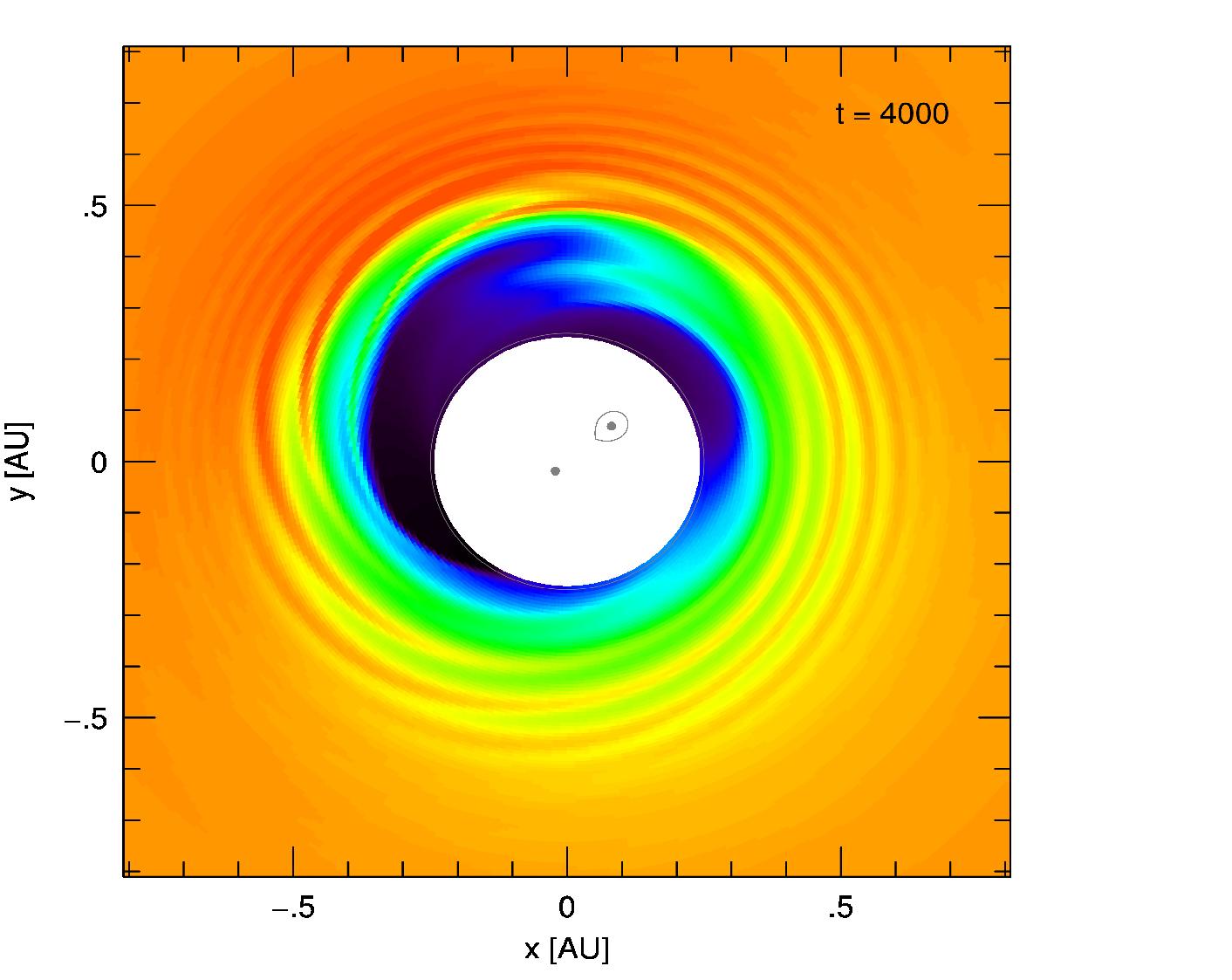} \\
\caption{Two dimensional density structure of an isothermal disk (model 1) around the central binary star.
The graph shows a local view around the binary. The computational grid extends from $r=0.25$ AU to $r=2.0$ AU.
The white inner region lies inside the computational domain and is not covered by the grid.
The positions of the primary and secondary are indicated by the gray dots. The the Roche-lobe of
the secondary is also shown.
}
\label{fig:2D-k38a}
\end{figure}

In Fig.~\ref{fig:2D-k38a}, we show a the two-dimensional surface density distribution for our isothermal
disk models. Note that the figure shows only the inner part of the computational domain around the central binary. 
The Roche lobe of the secondary star is shown as well. As shown here, an eccentric central binary strongly perturbs
the disk and produces time varying patterns. As indicated by the flat surface density profile, outside of the central 
gap, the disk shows less structure and is relatively homogeneous.

\begin{figure}
\center
\includegraphics[width=0.45\textwidth]{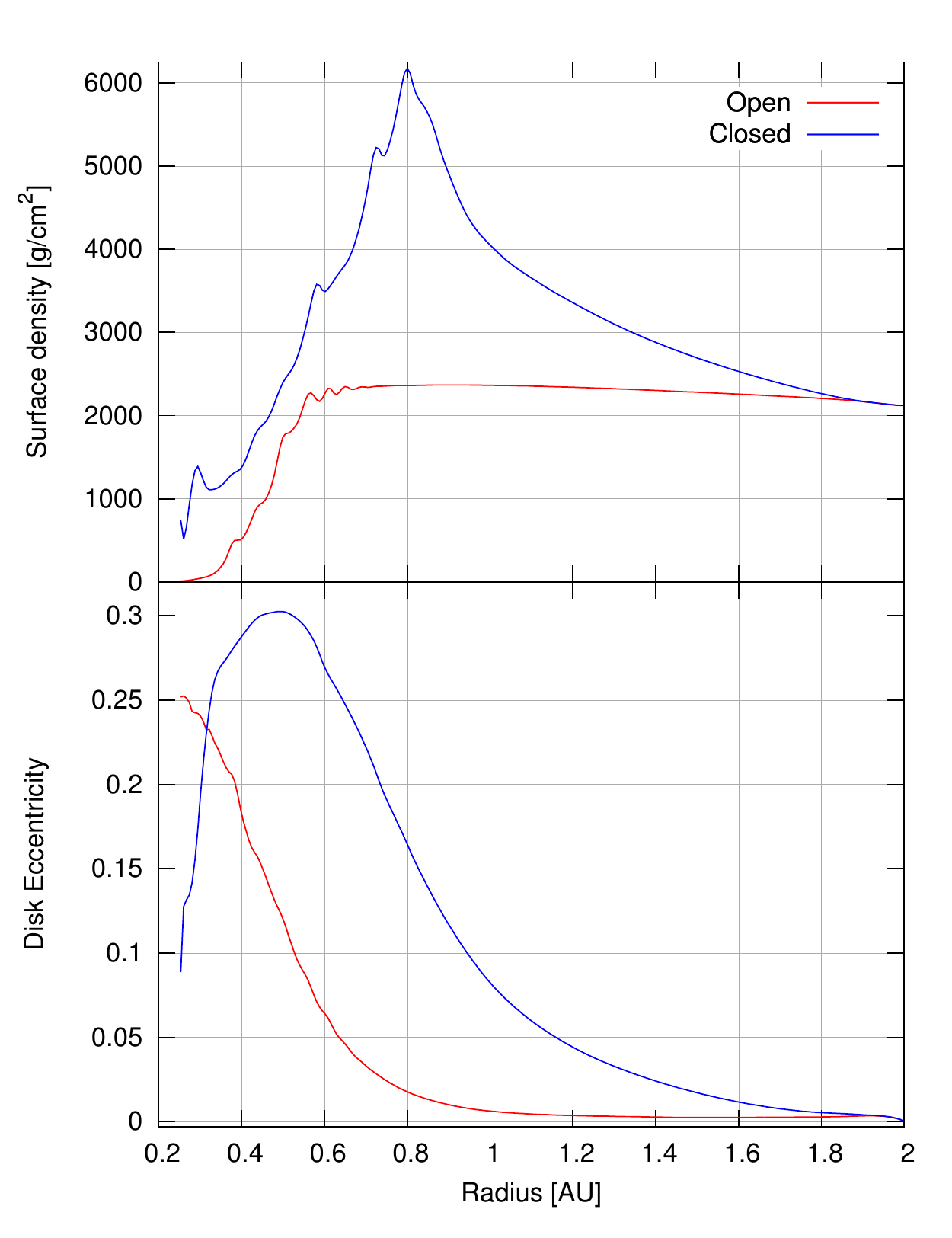} \\
\caption{Azimuthally averaged surface density and disk eccentricity profiles for two models with
different inner boundary conditions. The red curve corresponds to the standard model with an open inner boundary
and the blue curve represents a system with a reflecting inner boundary (model 2).
In both models, the disk is at equilibrium. 
}
\label{fig:closed-rmin}
\end{figure}

\subsection{A closed inner boundary}
\label{subsec:closed-rmin}
In this section, we study the effect of different boundary conditions at the inner edge of the computational
domain on the final results.
Inspired by the work of \citet{2013A&A...556A.134P}, who focused on models with closed inner boundaries, we
carried out simulations using their boundary conditions.
In Fig.~\ref{fig:closed-rmin}, we show the surface density and disk eccentricity for two simulations;
our standard model with an open inner boundary, and disk model 2 (see Table 1) with reflecting conditions 
at $r_{min}$ as used by \citet{2013A&A...556A.134P}.
Because in the latter model, material is not allowed to leave the disk through the inner boundary, it will 
accumulate near the outer edge of the gap, as seen by the spike in the density at $r = 0.8$AU in the top panel 
of Fig.~\ref{fig:closed-rmin}. At the same time, the eccentricity of the disk in this model becomes substantially 
larger than that in the model with the open boundary condition. Our results of the simulations with the closed inner 
boundary are very similar to those reported by \citet{2013A&A...556A.134P}, although these authors did not model 
the system of Kepler-38, but some related ones.

\begin{figure}
\center
\includegraphics[width=0.45\textwidth]{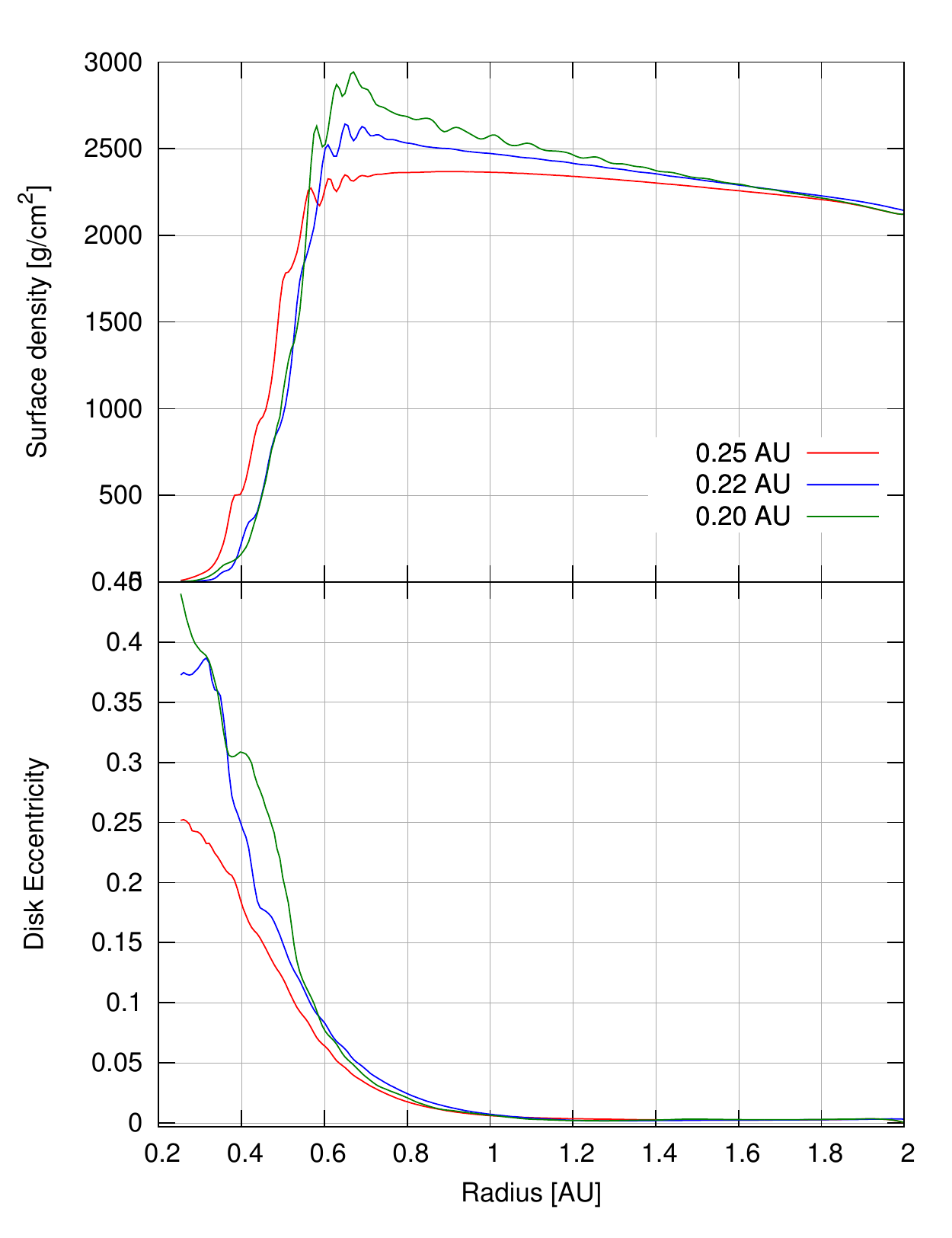} \\
\caption{Azimuthally averaged surface density and eccentricity profiles for three disk models with
different locations of the inner boundary, $r_{rmin}$. The red curve shows the standard model (model 1) with 
$r_{min} = 0.25$ AU, the blue curve is for a disk with $r_{min} = 0.22$ AU, and the green curve represents the
disk model with $r_{min} = 0.20$ AU (model 3). In all three models, the disk is at equilibrium. 
}
\label{fig:rminc}
\end{figure}

\subsection{Varying the location of the inner radius}
\label{subsec:rminc}
In this section, we examine the effect of the location of the inner boundary on the disk structure (model 3).
\citet{2013A&A...556A.134P} found that, at least in the case of a closed inner boundary, the amplitude
of the disk eccentricity depends on the location of $r_{min}$. In particular, if $r_{min}$ is chosen so large
that some MMRs between the binary and the disk fall outside the computational domain, 
the eccentricity of the disk will remain very small. To examine whether the disk will behave similarly for an open inner boundary,
we carried out two additional simulations with slightly smaller inner radii, $r_{min} = 0.22$ AU and $0.20$ AU.
Given the binary's semi-major, $a_\text{bin} = 0.1469$AU, these radii correspond to 1.5 and 1.36 $a_\text{bin}$,
respectively. 

As shown in Fig.~\ref{fig:rminc} (top panel), for small values of $r_{min}$, the density of the disk increases 
outside of the gap, and at the same time, the flow of the mass onto the binary is slightly reduced (to about $8 \times 10^{-7} M_\odot$/yr).
The profile of the edge of the gap, where the slope of the density is positive, is hardly affected by the choice of $r_{min}$.
However, the gap is slightly wider.
The eccentricity in the main part of the disk (outside $r=0.7$) does not vary with the location of $r_{min}$
(lower panel of Fig.~\ref{fig:rminc}). At $r=1$ the disk eccentricity is about $e_\text{disk}=0.1$
and beyond $r=1.0$ it is always below $0.01$.
Within the gap region, where the density is small, $e_\text{disk}$ rises to a maximum of $0.4$ 
at the inner boundary for the models with the smaller $r_{min}$. The disks with $r_{min} = 0.2$ AU and $r_{min} = 0.22$ AU show very similar
behaviour. For the sake of comparison, we also ran simulations with higher spatial resolution. The results were very similar to
those depicted by the two panels of figure 6.

Overall, results of our simulations indicate that in a model with an open inner boundary, 
the disk eccentricity $e_\text{disk}$ remains small in a large portion of the disk and independent of the location of the inner boundary.  
%
%
This is unlike the findings of \citet{2013A&A...556A.134P} who showed that in cases with close boundaries,
the disk eccentricity will rise to much larger values \citep{2013A&A...556A.134P}. The latter can be attributed 
to the fact that a reflecting inner boundary creates
a cavity which sustains the eccentric global mode of the disk. Our findings are also in agreement with
\citet{2008A&A...487..671K} who found an eccentricity reduction for circumstellar disks in close binary
systems when using outflow boundary conditions.
In our opinion, open inner boundaries are more realistic because they allow for the accretion of material onto the binary stars.
However, we would like to caution that in the case of more eccentric binaries, it has been found that an open inner boundary 
may cause the central cavity to be very large since disk material cannot re-enter the computational 
domain once it has been lost through $r_{min}$ (see \citet{2009A&A...508.1493M}).
This situation occurs primarily in highly eccentric binaries such as Kepler-34 which has an eccentricity of 0.52
(A.~Pierens, private communication).
For our system, Kepler-38, we found the opposite behaviour; a reduced disk eccentricity for an open inner boundary.
This follows from the fact that an open inner boundary reduces the strength of a global eccentric mode.
Because in this paper we are more interested in the evolution of planets in circumbinary disks, we do not study
this interesting behaviour of the disk any further and defer that to a future time.

We note that among the three models shown in figure 6, the model with the smaller inner radii can have a considerably 
smaller numerical time step. In addition to the smaller size of the grid cells, this effect is caused by the larger 
deviation from a uniform keplerian velocity as induced by the binary star, which makes the FARGO-algorithm less effective. 
Therefore, to save computational time and resources, in the remainder of this study, we will use for the inner
radius of the disk, a standard value of $r_{min} = 0.25$ AU, although we will present some comparison simulations using 
$r_{min} = 0.22$ AU as well.

\subsection{Planets in isothermal disks}
\label{subsec:iso-withplanet}

\begin{figure}
\center
\includegraphics[width=0.45\textwidth]{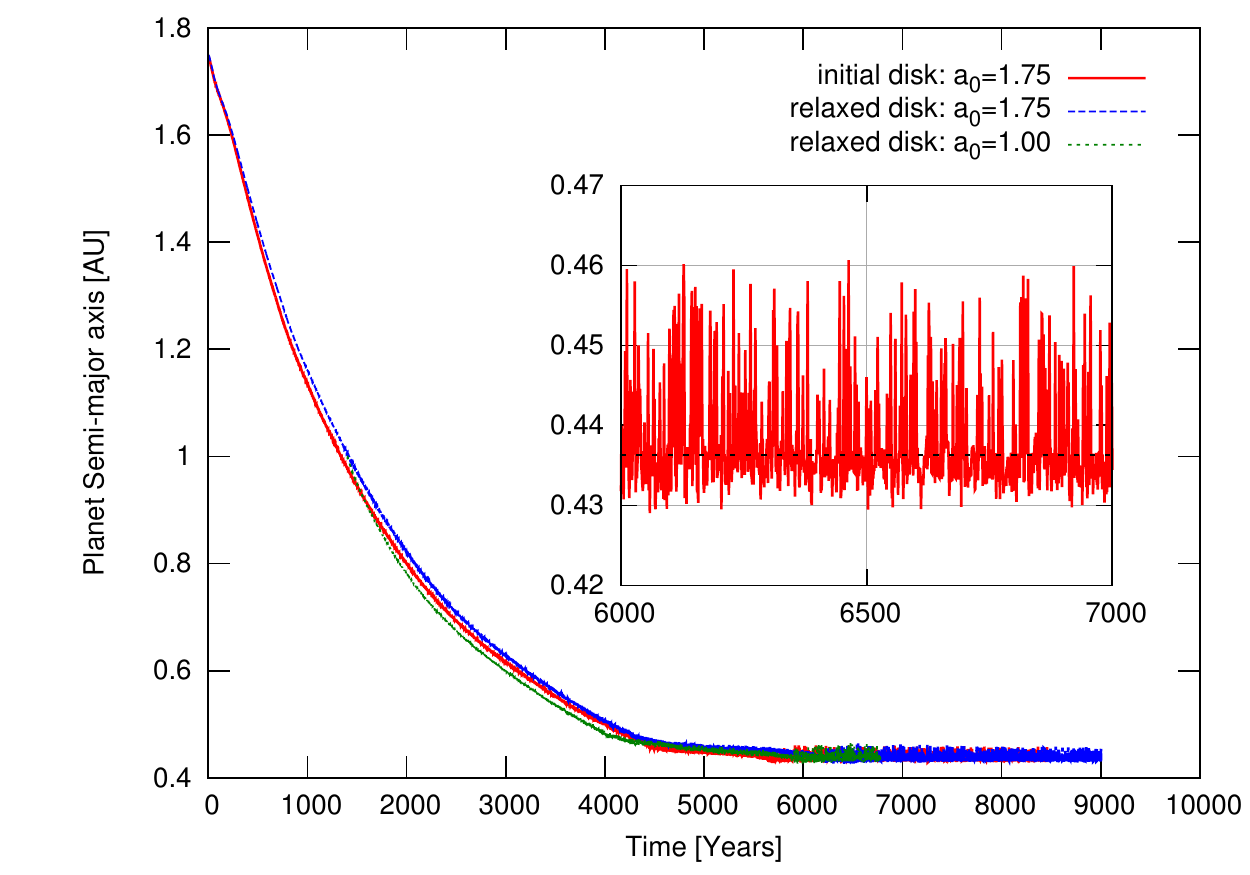} \\
\caption{The evolution of the semi-major axis of the planet in an isothermal disk (model 1).
 In two simulations, the planet is started in a circular orbit at a distance of $a_0 =  1.75$~AU 
from the center of mass of the central binary. 
In the simulation shown in red, the planet is included in the disk from the very beginning of the simulation
and it evolves as the disk evolves starting from its initial density profile.
In the simulation shown in blue, the disk is first relaxed to the equilibrium and then the planet is embedded
in it. In the simulation shown in green, the planet is started at $a_0 = 1.0$ AU in a relaxed disk.
The inset shows the results near the end of the simulation for the first model.
The dashed black line in the inset corresponds to the mean value of the planet semi-major axis at $a_\text{p} = 0.436$ AU.}
\label{fig:ap-k38a4a}
\end{figure}

\begin{figure}
\center
\includegraphics[width=0.45\textwidth]{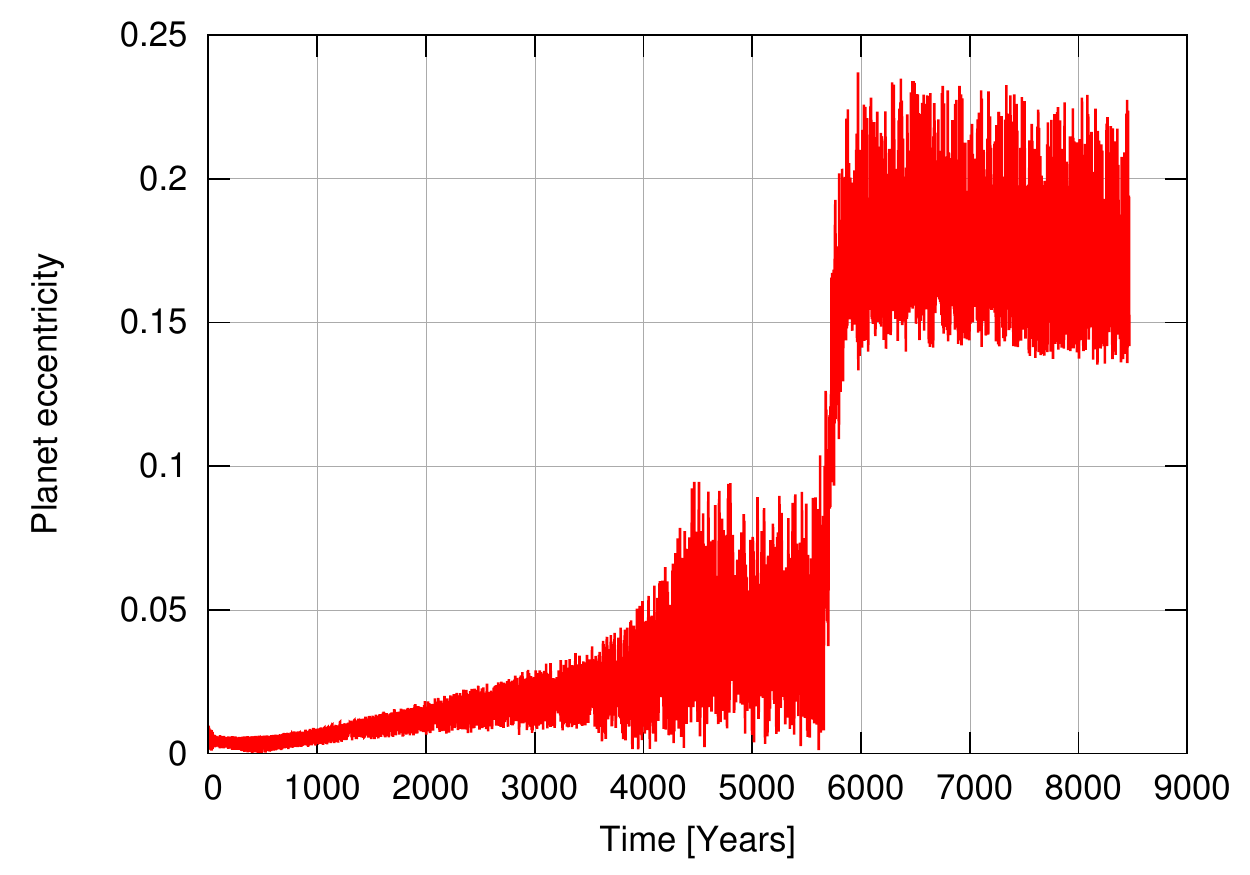}
\caption{Graph of the evolution of the eccentricity of the planet in an isothermal disk shown in red 
in Fig.~\ref{fig:ap-k38a4a} (model 1). The planet was embedded in the disk prior to the start of the disk evolution 
at a semi-major of $a_0 =  1.75$~AU. 
 }
\label{fig:ep1-k38a4a}
\end{figure}

To study the evolution of the system with an embedded planet, we start our simulations with a planet initially at different distances 
(semi-major axes, $a_0$) from the center of mass (barycenter) of the binary and in a circular orbit. 
The planet's evolution is determined by the gravitational action of the disk and the central binary.
As mentioned earlier, during the evolution of the planet, its orbital elements
are calculated using Jacobian coordinates with respect to the barycenter of the central binary.
The planet mass is fixed to 115 Earth masses, and there is no accretion onto the planet.
As a reference, we present in Tab.~\ref{tab:kepler38}, the orbital parameters of the planet of Kepler-38, as inferred from the observations.

Fig.~\ref{fig:ap-k38a4a} shows the evolution of planet through the disk.
We present here the results of three simulations that were carried out for different initial setups to examine the dependence
of the outcome on the initial state of the disk and planet’s orbital elements. In the first model (red line) the planet is 
started directly in the initial disk which has not been brought into an equilibrium (i.e. using the initial disk setup of model 1).
In this simulation, the disk and the planet evolve simultaneously.
In the other two simulations, the planet is started in a disk that is already in an equilibrium. 
The blue line corresponds to the planet starting at $a_0 = 1.75$ AU and the green line is when the planet starts at $a=1.0$ AU. 
We have shifted the last model in time to ensure an overlap with the first two.

As shown in the Fig.~\ref{fig:ap-k38a4a}, the semi-major axis of the planet continuously shrinks 
(i.e., the planet migrates toward the central binary) until it reaches the inner cavity created by the
binary. The strong drop in the disk surface density within the central cavity and the resulting positive density 
gradient act as a planet trap \citep{2006ApJ...642..478M}. Here the negative Lindblad torques are balanced
by positive corotation torques. As a result, the planet stops its inward migration in this region.
As shown in the figure, in all runs, the planet stops at approximately the same distance from the binary. 
The semi-major axis of the planet in the three runs has an average value of $a_\text{p} = 0.436$ AU
(blue line in inset of Fig.~\ref{fig:ap-k38a4a}). The results indicate that in all simulations, 
the system evolves to the same final state. At this state, the planet resides in an orbit with a period of about 100 days
which is slightly outside of its observed orbit (see Tab.~\ref{tab:kepler38}).
This implies that the outcome of the migration process does not depend on the history
of the system and is determined solely by the physical parameters of the disk. In an isothermal system and for a given binary, 
these parameters are determined by the values of $H/r$ and viscosity. 

Fig.~\ref{fig:ep1-k38a4a} shows the evolution of the eccentricity of the planet corresponding to the red model in Fig.~\ref{fig:ap-k38a4a}.
As shown here, during the planet’s inward migration, its eccentricity increases continuously until at $t=4500$ when it reaches a constant 
value close to 0.05 and oscillates around that value for approximately 1000 years. At time $(t \sim 5700)$, the planet is temporarily 
captured in a mean-motion resonance with the binary and its eccentricity undergoes a sudden jump \citep{1999ssd..book.....M}  
where it oscillates between 0.15 and 0.20. We will discussed this resonant capture in more detail in the next section.

\begin{figure}
\center
\includegraphics[width=0.49\textwidth]{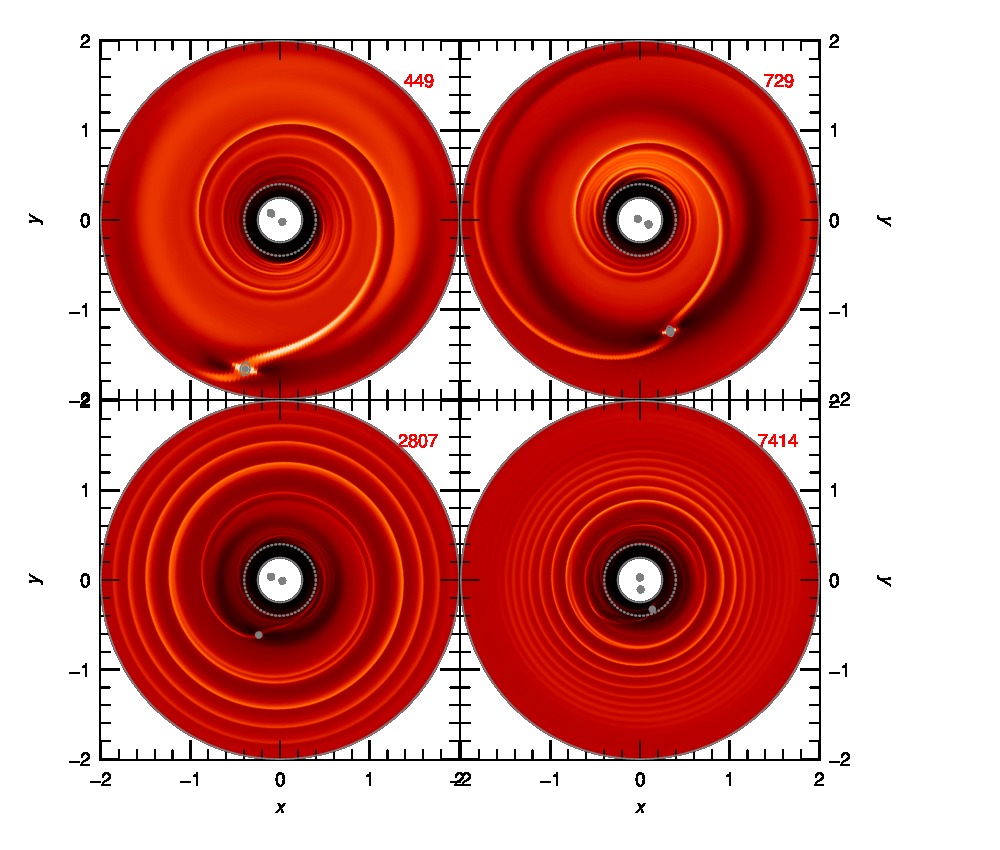}
\caption{Two dimensional snapshots of the evolution of the disk surface density with an embedded planet. 
The snapshots correspond to the times 449, 729, 2807, and 7414 years. The planet, shown as the gray dot with the 
Roche radius, was embedded in the disk at an initial distance of $r=1.75$ AU. The dashed gray line around the
central binary shows the approximate boundary of stability for planetary orbits.
A movie of the simulation shown here can be found in the supplementary online material.
 }
\label{fig:k38a5a-multi}
\end{figure}

\begin{figure}
\center
\includegraphics[width=0.45\textwidth]{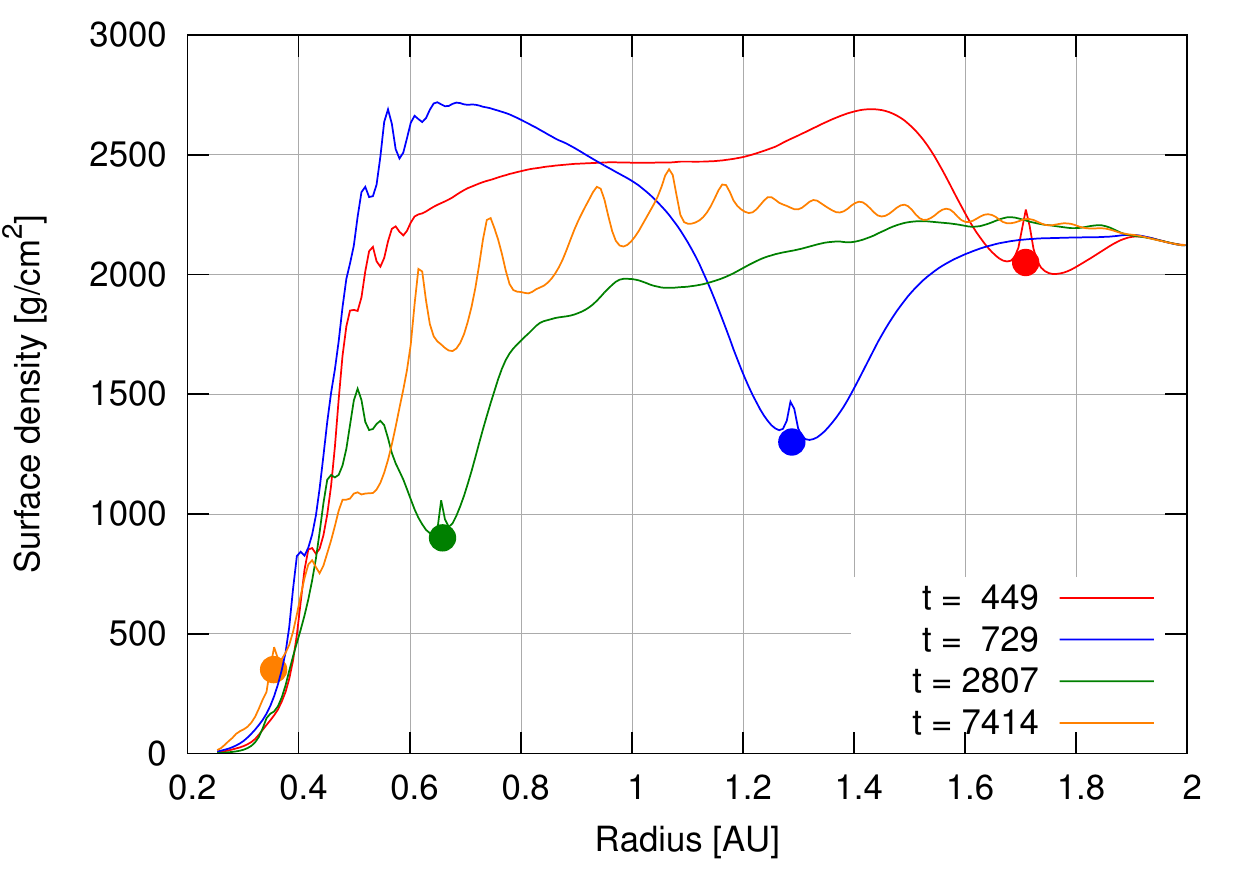}
\caption{Graph of the averaged radial surface density of a disk with an embedded planet. The colors correspond
to the times of the snapshots shown in Fig.~\ref{fig:k38a5a-multi}. The big circle in each graph represents the radial 
position of the planet at the time corresponding to the  graph.
For illustrative purposes, we have moved the circles close to their corresponding curves.
 }
\label{fig:k38a5a-sig}
\end{figure}

Fig.~\ref{fig:k38a5a-multi} shows four two-dimensional snapshots of the evolution of the disk surface density 
during the inward migration of an embedded planet in one of our models. One can clearly see the spiral disturbances created 
by the planet during its migration. The last panel, corresponding to $t=7417$, shows the final position of the 
planet in an orbit with a semi-major axis of $a_\text{p} = 0.436$ AU (see Fig.~\ref{fig:ap-k38a4a}).
In Fig.~\ref{fig:k38a5a-sig}, we plot the azimuthally averaged surface density of the disk corresponding to the
same times as in the snapshots of Fig.~\ref{fig:k38a5a-multi}. The colored circles in this figure represent the planet
in its actual radial distance from the center of mass of the binary.
A planet of this mass does not produce a full gap, but only clears out about a third of the surface
density. As the planet moves inward, the surface density of the inner part of the disk is strongly distorted.
However, it regains its unperturbed profile after the planet has reached its final position,  
(see Fig.~\ref{fig:k38a-sigecc}).

\begin{figure}
\center
\includegraphics[width=0.45\textwidth]{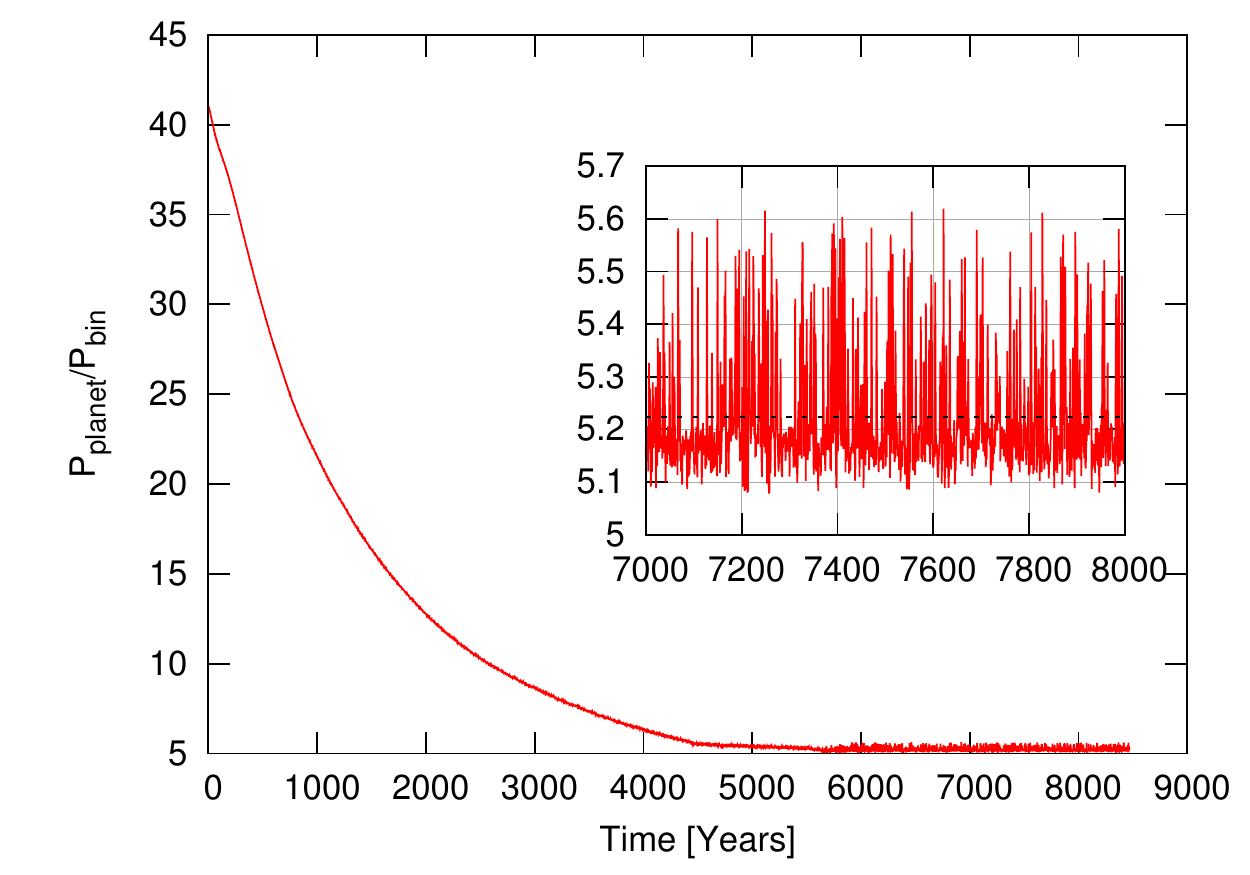}
\caption{The evolution of the period ratio of the planet and binary for
the model where the planet is embedded in the initial disk at $a_0 =  1.75$~AU. 
The inset shows the period ratio for the final stage of its evolution. The dashed line in the inset
corresponds to the mean value of $P_\text{planet}/P_\text{bin} = 5.22$.
 }
\label{fig:k38a4-prat1-multi}
\end{figure}

\begin{figure}
\center
\includegraphics[width=0.45\textwidth]{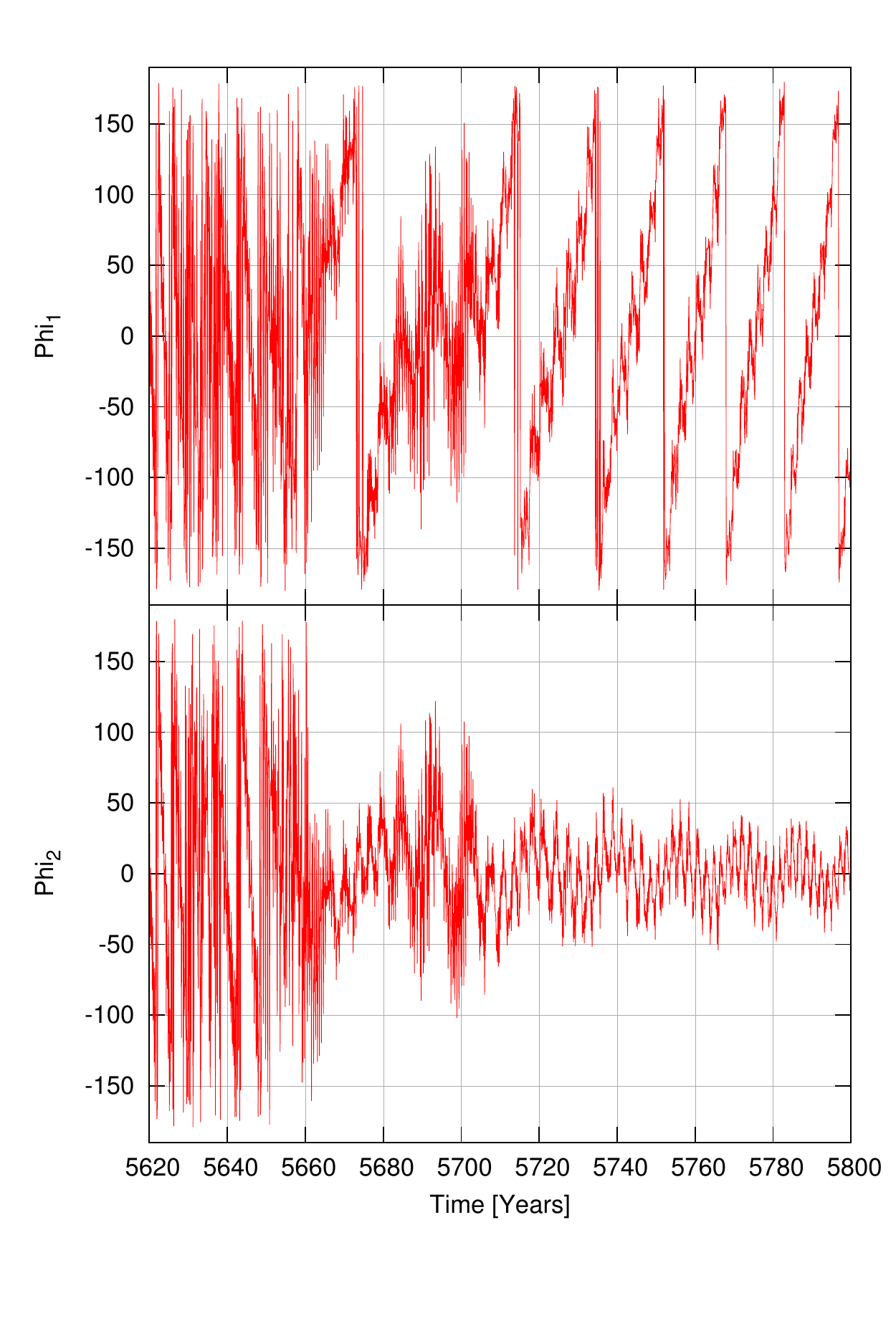}
\caption{Graphs of the resonant angles $\Phi_1$ and $\Phi_2$ (see Eqs.~\ref{eq:phi1} and \ref{eq:phi2})
during the capture phase in a 5:1 mean-motion resonance.
 }
\label{fig:k38a4a-phim}
\end{figure}

\subsubsection{Capture in Resonance}
The evolution of the semi-major axis of the planet and in particular, the sudden increase in its eccentricity 
at time $t \approx 5700$ seem to point to a temporary capture in a mean-motion resonance with the central binary. 
To further analyze this, we plotted the variations of the period-ratio of the planet and binary,
$P_\text{planet}/P_\text{bin}$, during the inward migration of the planet. Fig.~\ref{fig:k38a4-prat1-multi} shows
the results. As shown here, as the planet migrates inward, this period-ratio continuously decreases until it reaches
a constant value at $t \approx 5000$. The inset in Fig.~\ref{fig:k38a4-prat1-multi} shows the evolution of the system 
near the end of the simulation for one thousand years. The black dashed line in this inset denotes the arithmetic 
averaged value of the period ratio, $P_\text{planet}/P_\text{bin} = 5.22$, suggesting a possible capture into a 5:1 
MMR with the binary.

To determine whether the planet is truly captured in a 5:1 MMR, its corresponding resonant angles have to be analyzed.
The capture in a resonance between two bodies occurs when 
at least one of the resonant angles of the system librates around a certain value and does not cover the full range of
0 to 2$\pi$ (see \citet{1999ssd..book.....M} for details). For a general $p$:$q$ commensurability with $p>q$ 
and in a planar system, the resonant angles are defined as
\begin{equation}
\label{eq:phik}
   \Phi_{k} = p \lambda_\text{p} - q \lambda_\text{b} - p \varpi_\text{p} + q \varpi_\text{b}
     + k (\varpi_\text{p} - \varpi_\text{b}),
\end{equation}
where $\lambda_\text{b}, \lambda_\text{p}, \varpi_\text{b}$ and $\varpi_\text{p}$ denote the mean longitudes
and longitude of periapse for the inner binary (b) and outer planet (p). The integer $k$ in this equation 
satisfies the condition $q \leq k \leq p$ and has $p-q+1$ possible values. Of the $p-q+1$ resonant angles, at most two 
are linearly independent. That means, in a actual resonant configuration, at least one of these angle will librate \citep{2002MNRAS.333L..26N}.
As our numerical simulations indicate the possibility of a capture in the 5:1 MMR, we studied the time-variations of all resonant angles 
of our system. Note that in this case $p=5$, $q=1$, and $k$ runs from $1$ to $5$. 
Fig.~\ref{fig:k38a4a-phim} shows the evolution of the following two resonant angles for 180 yrs from $t=5620$ to $t=5800$,
\begin{equation}
\label{eq:phi1}
  \Phi_1  = 5  \lambda_\text{p}  - \lambda_\text{b} - 4 \varpi_\text{p} \,,
\end{equation}
and
\begin{equation}
\label{eq:phi2}
  \Phi_2  = 5  \lambda_\text{p}  - \lambda_\text{b} - 3 \varpi_\text{p} - \varpi_\text{b} \,.
\end{equation}
As shown here, before $t=5670$, the evolution of these angles is rather irregular confirming that the planet was not in a resonance.
After this time, $\Phi_1$ shows a regular circulating behaviour, whereas the angle $\Phi_2$ begins to librate around zero with 
an amplitude of less than 40 degrees. This librating evolution of $\Phi_2$ indicates that the planet is indeed captured in a
5:1 MMR with the binary star. Our analysis of all other resonant angles of the system, $\Phi_3$, $\Phi_4$, and $\Phi_5$ showed that
these all have circulating behaviours as well, sometimes prograde (as $\Phi_1$) and sometimes retrograde.
For all cases shown in Fig.~\ref{fig:ap-k38a4a}, the planet remains captured in the 5:1 resonance for the duration of the simulations.
We also studied the behaviour of the longitude of periastron of the binary star, $\varpi_\text{bin}$, during the slow inward migration 
of the planet. Our analysis indicated that prior to the capture of the planet in a 5:1 MMR, $\varpi_\text{bin}$ was precessing in a prograde 
sense. After capture into the 5:1 resonance, this precession became retrograde.

\noindent

\begin{figure}
\center
\includegraphics[width=0.45\textwidth]{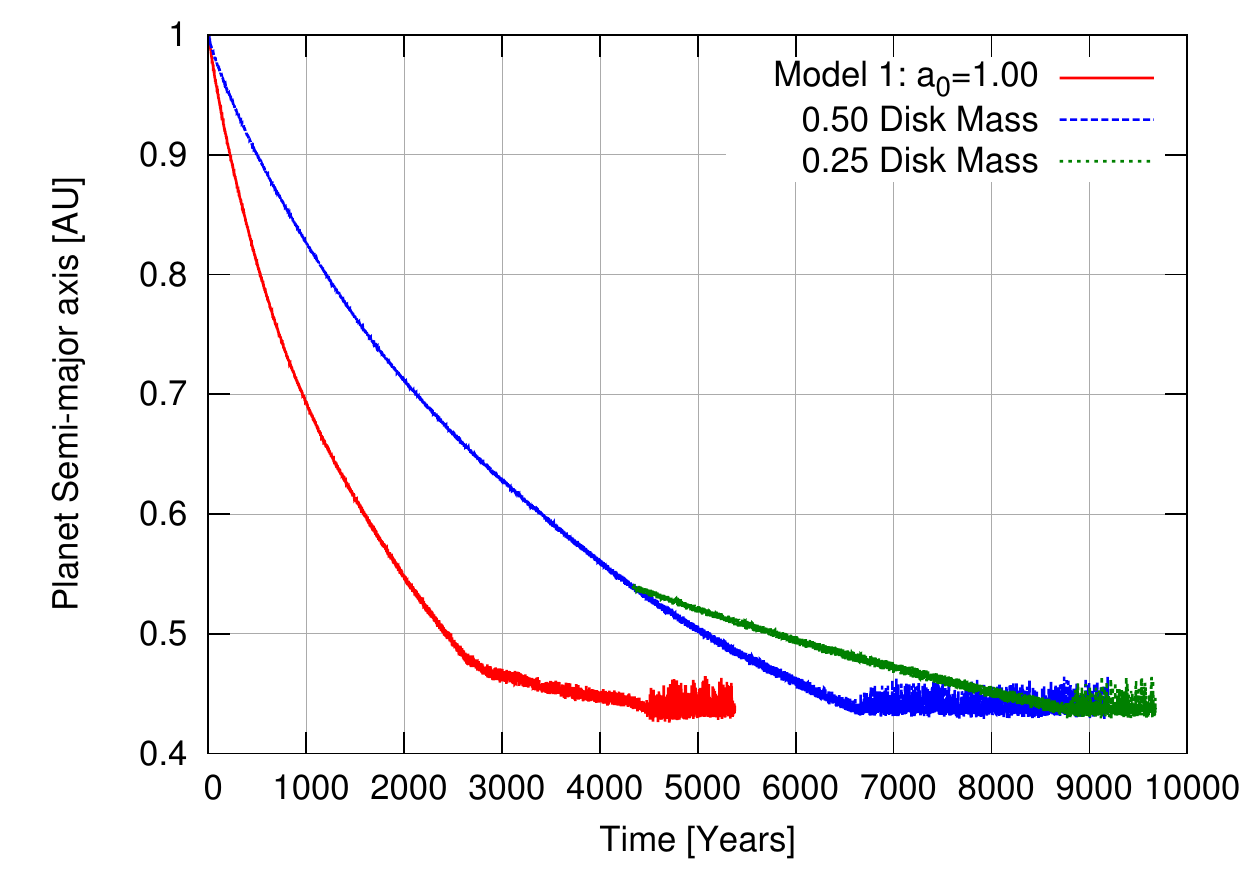}
\caption{The evolution of the semi-major axis of a planet embedded in an isothermal disk 
for different values of the disk mass. The graph in red corresponds the simulation of the standard disk model with the planet 
initially starting at $a_0=1.0$ AU. The graph in blue shows a similar simulation but with half the disk mass. The graph in green
corresponds to a simulation with a quarter of the initial disk mass and continued from the model in blue at $t=4300$. 
 }
\label{fig:k38ma4-apc1}
\end{figure}
\subsubsection{Variation of Disk mass}
Up to this point, the presented results and analyses were based on simulations in which, as shown in Fig.~\ref{fig:k38a-diskmass},
the disk had a given mass. To examine whether the results will depend on the mass of the disk, we carried out simulations 
where the disk surface density was reduced by a factor of two and four. 
Results indicated that disks with reduced surface density show similar density evolution as those shown by the top panel of
Fig.~\ref{fig:k38a-sigecc}, but only scaled down by an appropriate factor. This identical density evolution can be attributed 
to the fact that in an isothermal disk with an $\alpha$-type viscosity and no self-gravity,
the surface density of the disk is scaled out of the normalized equations of state making the disk evolution independent of the actual density. 

In a disk with a smaller mass and surface density, the inward migration of the planet will be slower. 
Fig.~\ref{fig:k38ma4-apc1} shows the evolution of a planet in an isothermal disk with different masses. 
The red line in this figure represents the inward migration of a planet started at $a_0=1.0$ AU in our standard disk model.
The blue line corresponds to the migration of the same planet in a disk with half the mass of our standard model. 
As shown here and as expected, the inward migration is about two times slower than in the standard case. 
We started our third simulations (green line) with a quarter of the disk mass and to save computational time, continued it 
from the 2nd mode (blue line) at time 4300. The planet migration rate is again reduced by a factor of two. 
As pointed out, for instance, by \citet{2013A&A...556A.134P} when analyzing the system of Kepler-16,
in a system with a slower rate of migration, a capture in higher order MMRs is more expected.
These authors found that Kepler-16b could have undergone a temporary capture in a 6:1 MMR with the central binary.
In our simulations, the planet is always captured in a 5:1 resonance and in the same configuration, where
only $\Phi_2$ librates. The eccentricity of the disk also reaches the same value in all three cases.

\begin{figure}
\center
\includegraphics[width=0.45\textwidth]{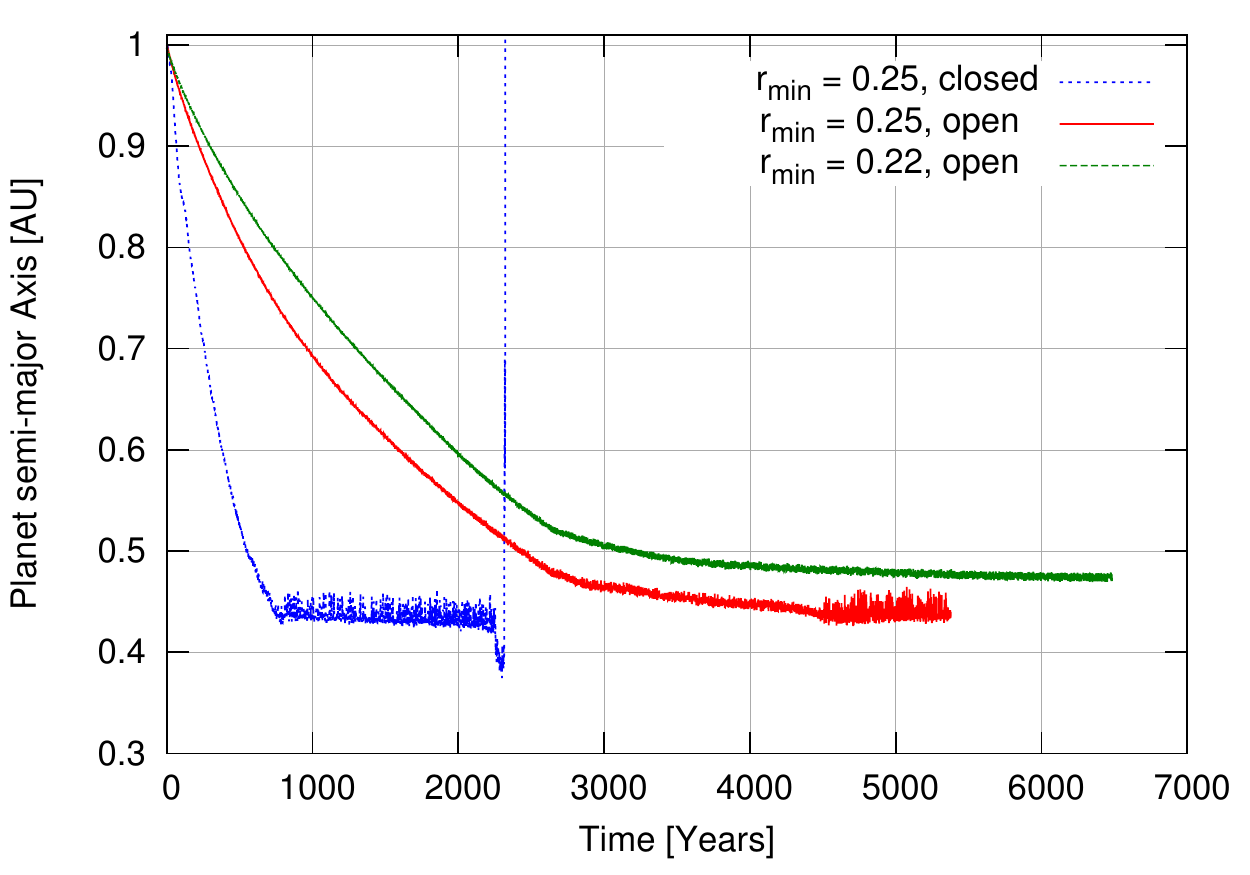}
\caption{The evolution of the semi-major axis of a planet embedded in an isothermal disk for different locations of
the disk inner boundary. The graph in red corresponds to the standard model with an open inner boundary at 0.25 AU.
The graph in blue represents a model with a closed inner boundary at 0.25 AU, and the graph in green is for a disk
with an open inner boundary at 0.22 AU.
}
\label{fig:k38za4-apc}
\end{figure}

\subsubsection{Planets in disks with different inner boundary conditions at $r_{min}$}
To test the sensitivity of our simulations to the choice of the disk boundary conditions at $r_{min}$, we ran
three different models. In two models we considered ${r_{min}}=0.25$ AU and once we assumed the inner boundary to be closed 
to mass in-flow, and once we considered it to be open (see model 2 in section ~\ref{subsec:closed-rmin}). 
We also ran a third model for which we moved $r_{min}$ to a different 
inner radius at 0.22 AU (see model 3 in section ~\ref{subsec:rminc}).
Fig.~\ref{fig:k38za4-apc} shows the evolution of the planet’s semi-major axis for these three cases. The planet was started
at $a_0=1.0$ AU in an already relaxed disk in all three cases. See
Figs.~\ref{fig:closed-rmin} and \ref{fig:rminc} for the density and eccentricity distribution of models 2 and 3.

Due to the higher density in the simulation with a closed inner boundary (shown in blue), the planet migrates inward at a 
much faster speed. Similar to the previous simulations, in this case the planet is captured in a 5:1 MMR with the binary. 
At this state, the planet’s resonance angle $\Phi_2$ began to librate around zero and its other resonance angles start circulating.
In contrast to the previous simulations, the fast inward migration of the planet causes its orbital eccentricity to be slightly higher
($e_\text{p} = 0.25$) than in the case when the disk had an open inner boundary. This larger value of eccentricity leads to orbital
instability as a result of which the planet leaves the resonance at time $t \approx 2300$ and is scattered out into an orbit with 
much larger semi-major axis and very high eccentricity.

In the simulations where the open inner boundary has been shifted to $r_{min} = 0.22 AU$,
the planet migrates inward at a reduced speed and reaches $a_\text{p}=0.47$ AU at the
end of the simulation. This is in an excellent agreement with the observed orbit of Kepler-38b. 
The eccentricity of the planet at this state oscillates between 0.01 and 0.06 with an average around 0.03.
These results confirm the expectation that the final position of the planet is determined by the
location of the outer edge of the central cavity. 

In summary, in the model in which the outer edge of the central cavity is at a farther distance and has the unrealistic
boundary condition of being closed to mass in-flow, the planet crosses the 5:1 MMR and eventually becomes unstable.
When the outer edge of the central cavity is at a closer distance (e.g., 0.22 AU) and has the more realistic boundary
condition of being open to mass in-flow, the planet stops its inward migration outside of the 5:1 MMR and maintain its
orbital stability for a long time. We will return to this point when we discuss radiative models in the next section.


\begin{figure}
\center
\includegraphics[width=0.45\textwidth]{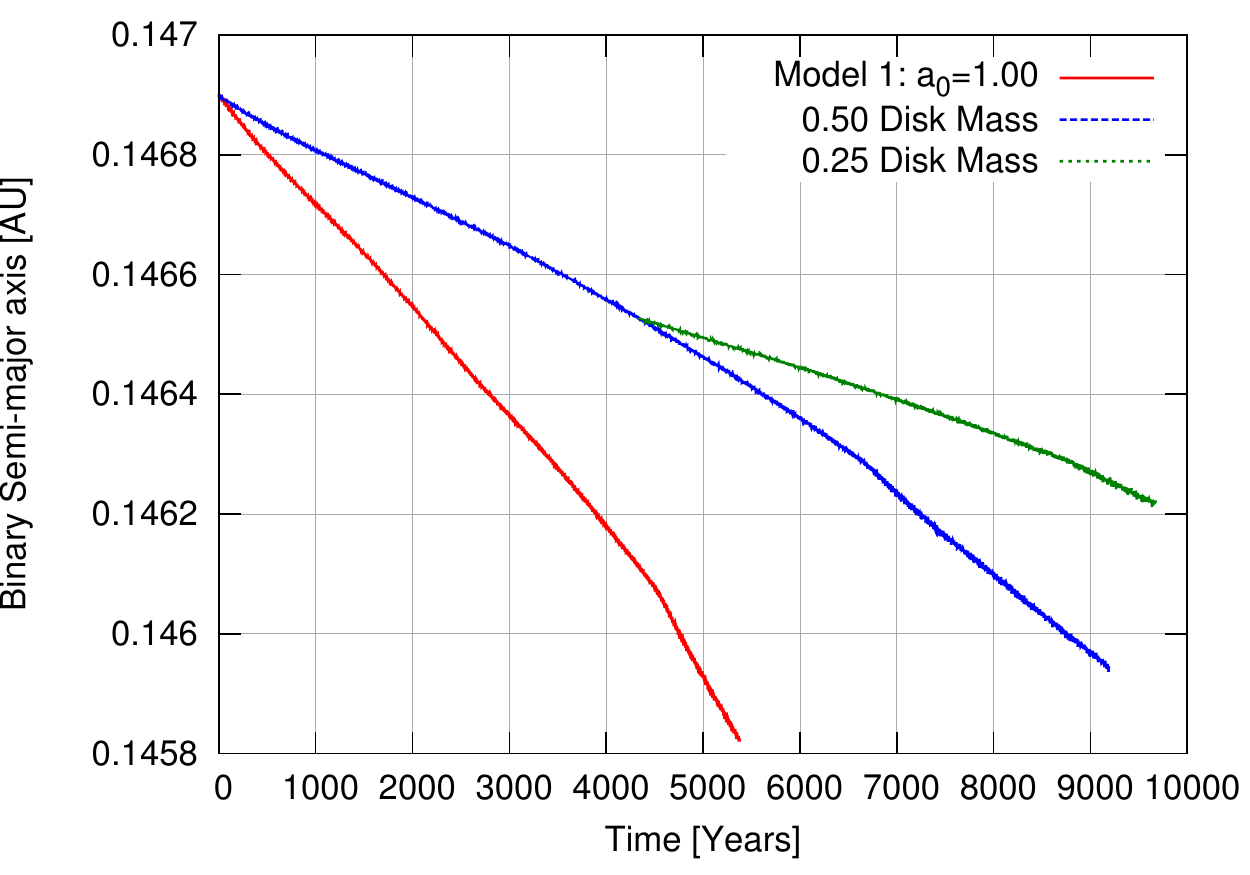}
\caption{The evolution of the semi-major axis of the binary in a system with an isothermal disk for different values of the
disk mass. The graphs uses the same models and colors as in Fig.~\ref{fig:k38ma4-apc1} for the evolution of the planets.
 }
\label{fig:k38ma4-abc1}
\end{figure}

\subsubsection{The orbital elements of the binary}
As mentioned earlier in the description of our models and their setups, the motions of all objects (including the stars of the binary) 
are affected by the gravity of the disk. As a result, the orbit of the central binary will change in time as it interacts with the disk 
and the planet. In Fig.~\ref{fig:k38ma4-abc1}, we plot the evolution of the semi-major axis of the central binary in three systems
with disks with masses as in Fig.~\ref{fig:k38ma4-apc1}. One can see from this figure that during the evolution of the system, the 
semi-major axis of the binary reduces with time. This can be explained noting that while the disk interacts with the binary, angular 
momentum is transferred from the binary to the disk material causing the orbit of the binary to shrink.
As expected, disks with smaller masses lead to a slower decline in the orbit of the binary. At the point where
the planet is captured into resonance, the binary loses angular momentum more rapidly.

%

\section{The radiative case}
\begin{figure}
\center
\includegraphics[width=0.45\textwidth]{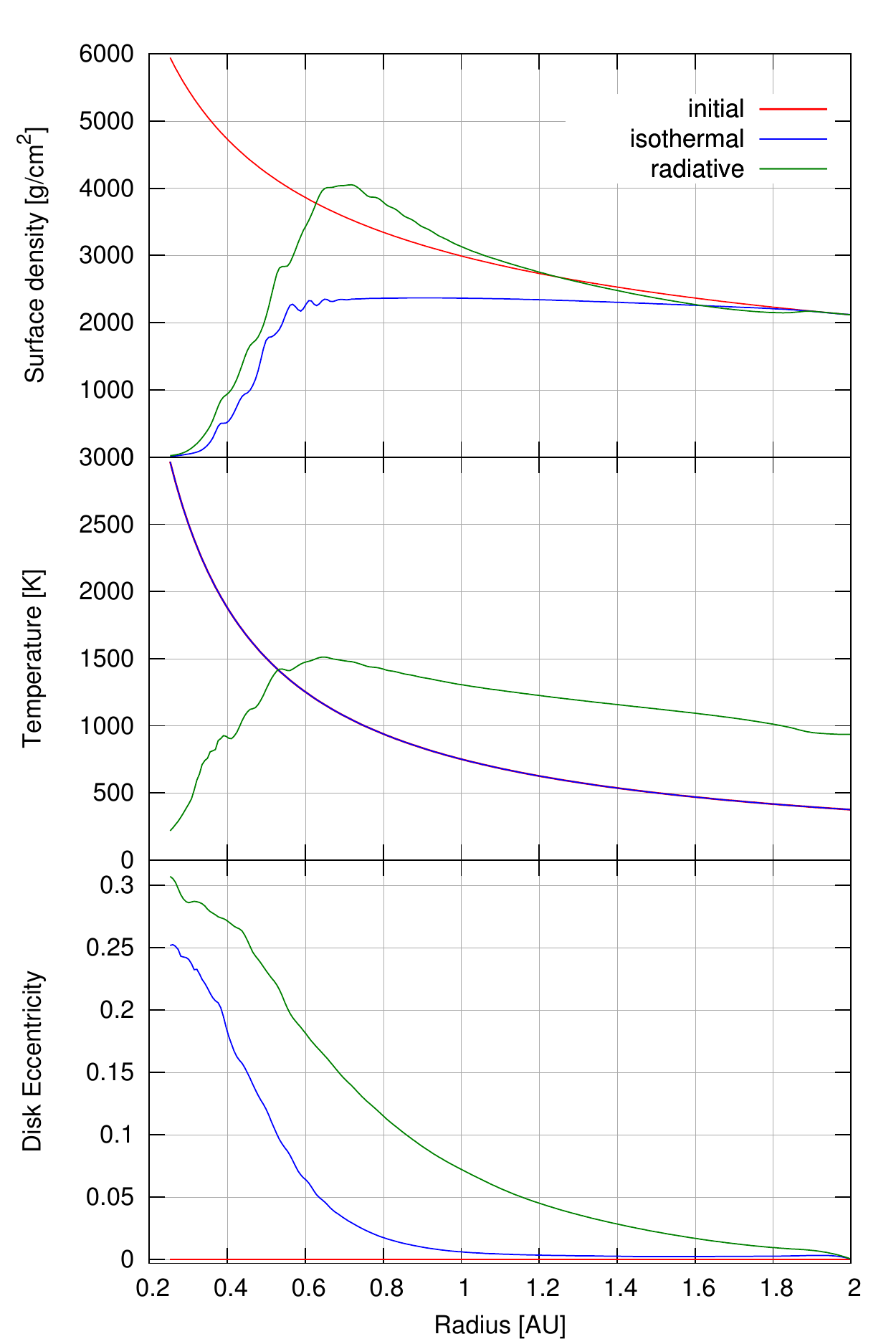} \\
\caption{Graphs of the surface density (in g/cm$^2$, top panel), temperature (in K, middle panel) and
eccentricity (bottom panel) of a radiative disk in the initial model (red), an isothermal case with $H/r=0.05$
(model 1, blue), and a quasi-equilibrium state (green).
 }
\label{fig:k38f-structure}
\end{figure}

As mentioned earlier, although the study of a locally isothermal disk can portray a general picture
of the evolution of the system and its planet, it does not present a realistic model. For instance,
in an isothermal disk, despite the strong heating action of the spiral waves that are induced by the 
binary star, the temperature does not vary and is held constant. However, because the migration of 
the planet and the evolution of its eccentricity depend on the disk temperature \citep{1988Icar...73..330W},
a more realistic model should allow the disk temperature to vary, as well. In this section, we consider the latter.
Because the evolved quantity in the simulations is the midplane temperature,
the inclusion of an energy equation with viscous dissipation, radiative cooling from the disk surfaces, and diffusion 
in the disk's midplane will result in a self-consistent temperature determination in the disk, and make the models 
much more realistic.

Similar to the case of isothermal disks, we first study the evolution of the disk without a planet, 
and then evolve the planet in the disk. The models are generally started from the same initial conditions as in the 
isothermal ones. They are then evolved to an equilibrium state which now takes about 1500 years to reach.
We have compared the results to models which were started from the equilibrated isothermal states and found that
they both reach to identical final configurations. There was no gain in computational time when started from an equilibrium 
isothermal disk because the timescale for the disk adjustment was identical as it is determined by the radiative 
and viscous diffusion timescale.

\subsection{Disk structure}
Fig.~\ref{fig:k38f-structure} shows a comparison between the density and temperature of a radiative disk and that of
an isothermal model after they have reached their corresponding equilibrium states. 
The boundary conditions for the disk density are the same in both models. That means, the density near the outer boundary
is relaxed to the initial value which is held constant. This implies that the density for both models is identical near the outer 
boundary at $r=2.0$AU. In the inner parts of the disk, however, the radiative model has a higher density throughout 
(top panel in Fig.~\ref{fig:k38f-structure}). This high value of density may be due to the larger eccentricity of the disk 
(bottom panel in Fig.~\ref{fig:k38f-structure}), similar to the evolution of an isothermal disk with the closed inner boundary.
The total mass of the radiative model in the computational domain is approximately 18\% higher than that in
isothermal cases. The oscillations in the disk mass in the quasi-equilibrium state reach  $\sim 1.5\%$ which
is about 10 times larger than those in the isothermal models (see Fig.~\ref{fig:k38a-diskmass}). These larger variations 
are caused by the larger disk eccentricity in the radiative models, which results in
large variations in the mass flow rate through the disk, with an equilibrium value of
$\dot{M}_\text{disk} \approx 1.2 \times 10^{-6} M_\odot$/yr. Because of higher temperatures in the disk, in these models, 
the inner edge of the disk is slightly closer to the binary, compared to the isothermal cases.

As shown in the middle panel of Fig.~\ref{fig:k38f-structure}, the viscously heated disk is much hotter
in the regions outside the inner cavity than an isothermal disk with the same initial profile.
The increased temperature of the disk results in a larger vertical height of $H/r \approx 0.06$ to $0.075$ 
within the main part of the disk, compared to $H/r = 0.05$ for the locally isothermal disk.
Inside the cavity in the radiative model, the temperature becomes very low because the reduced density leads to optically
thin disks that can cool more efficiently. Also, the diffusion of radiation in the midplane of the
disk leads to additional loss of energy through the open inner boundary (stellar irradiation was not included in these models).

While the disk eccentricity in isothermal models was small, we find that in 
radiative models, the disk eccentricity becomes considerably large (see bottom panel of Fig~\ref{fig:k38f-structure}). 
This large eccentricity is caused by the higher disk temperature in the radiative case. 
For disks with less mass, the radiative models have smaller temperatures and lower eccentricities,
(see Sect.~\ref{subsec:rad_lowmass} below). The high eccentricity of the disk in radiative models
is in agreement with the results of \citet{2013A&A...556A.134P}, who also consistently find higher disk eccentricities for
larger $H/r$. Surprisingly, this behaviour is in contrast to the case where the disk surrounds one star of a binary.
In those cases, disks with smaller $H/r$ tend to have larger eccentricities \citep{2012A&A...539A..18M}. We do not have a full
explanation for this behaviour, but we suspect that the tidal truncation of the disk in this case, along with the 
eccentricity of the binary may play a role. In circumbinary disks, there is no outer radius that may handicap 
the growth of large eccentric modes.

\begin{figure}
\center
\includegraphics[width=0.45\textwidth]{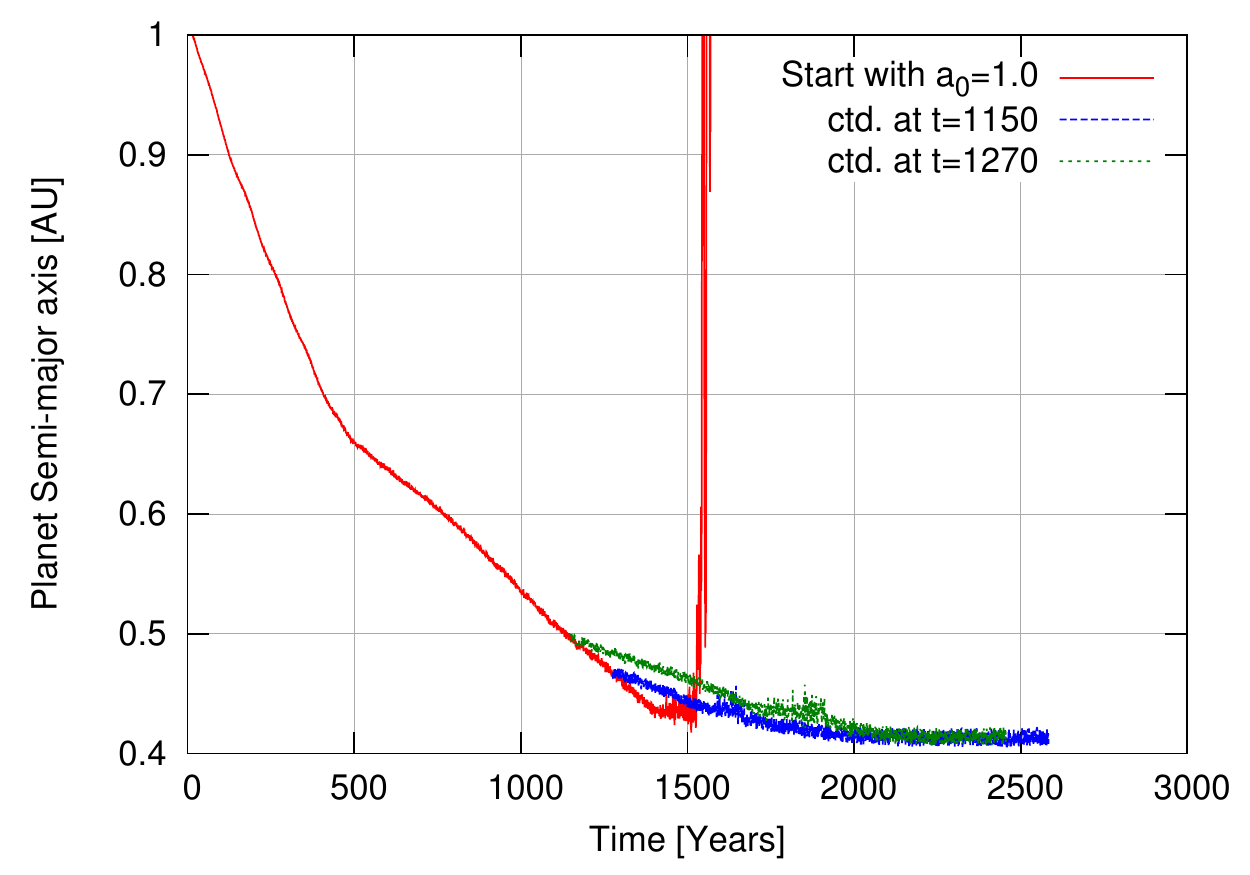} \\
\includegraphics[width=0.45\textwidth]{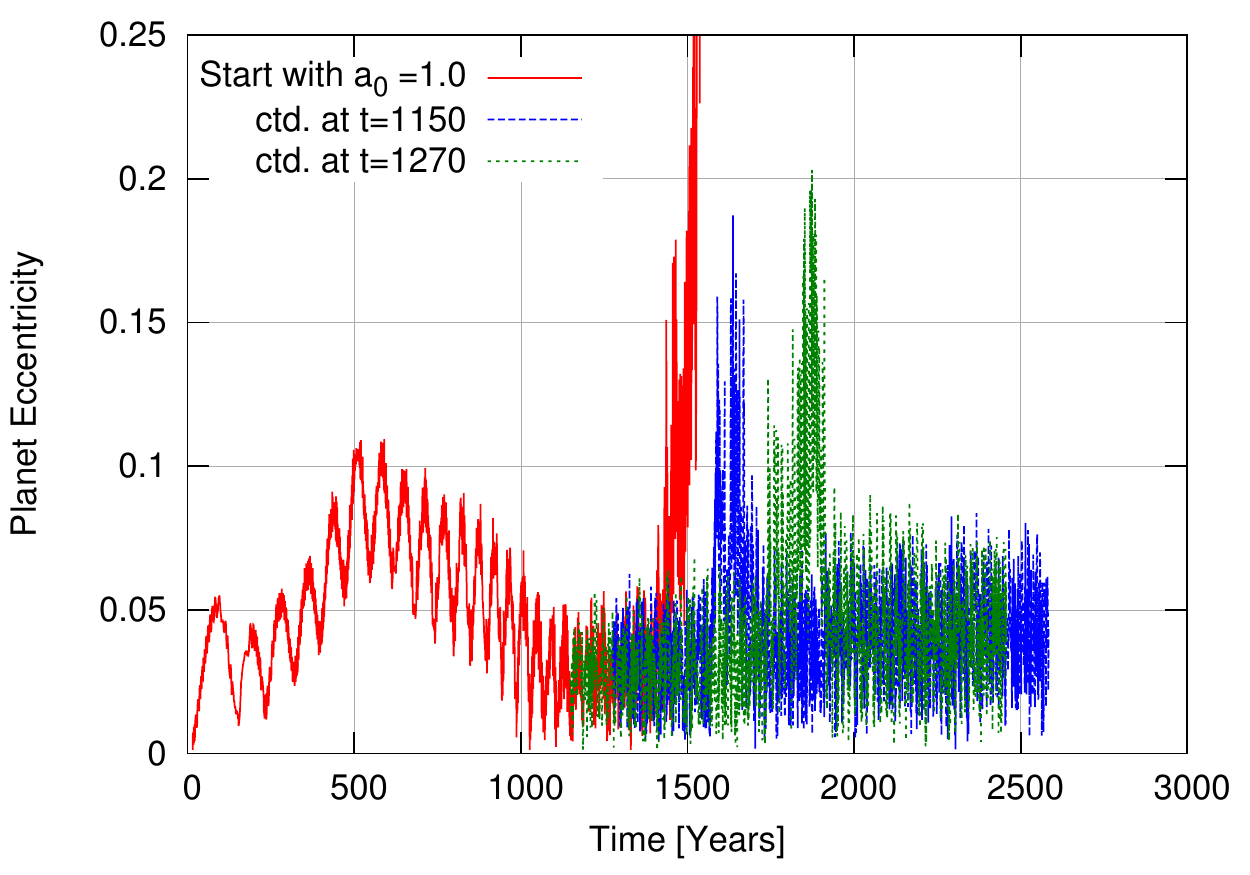}
\caption{Evolution of the semi-major axis and eccentricity of an embedded planet in a radiative disk.
The planet was started at $a_0 = 1.0$ AU (red curve). In the two other models shown in green and blue,
simulations were continued at specified times with slightly different numerical parameter (see text for details).
 }
\label{fig:k38f-planet}
\end{figure}

\subsection{Planets in radiative disks}

Similar to the isothermal case, after the radiative disk has reached thermal and dynamical quasi-static equilibrium, we place 
a $m_\text{p} = 0.34 M_\text{Jup}$ planet in the disk and follow its evolution. Because in previous models, the final outcome 
did not depend on the planet's initial condition, and also because the evolution of the planet is driven by dissipative processes, 
we expect that in the radiative case, the final outcome will also be independent of the initial condition, and 
we place the planets in a circular orbit with a semi-major axis of $a_0 = 1.0$AU. This choice of semi-major axis will allows us 
to save considerable computer time. 

Fig.~\ref{fig:k38f-planet} shows the evolution of the semi-major axis and eccentricity of the planet.
As shown here, because of the higher disk density in the radiative case, the planet migrates inward at a faster pace than 
in the corresponding isothermal model (see Fig.~\ref{fig:ap-k38a4a}). Similar to the isothermal case, the inward
migration slows down when the planet reaches the central cavity with the positive density slope. In the radiative case, this
occurs at $t \approx 500$ years. After this time, the migration continues and the planet reaches the location of 5:1 MMR 
with the binary at $t \approx 1400$ years where it is captured in that orbit. As expected, upon reaching the resonance, the planet 
eccentricity increases sharply (bottom panel). Shortly after capture in resonance at about $t=1500$ year, the planet’s orbit 
becomes unstable and the planet is scattered into a high eccentric orbit with a much larger semi-major axis.

To examine the effects of the numerics on the results, we changed two numerical parameters that can influence the calculations 
of the gravitational force between planet and the disk, namely the gravitational smoothing parameter $\epsilon$ (from 0.6 as in the
isothermal case to 0.7) and the force truncation parameter $p$ (from 0.6 to 0.8). We then carried out simulations for the evolution 
of the disk and planet. The two runs with these new parameters were continued from the first model at $t=1150$ and $t=1270$ years 
(green and blue lines in Fig.~\ref{fig:k38f-planet}), respectively, shortly before instability set in. In both models, the planet 
is temporarily captured in a 5:1 resonance with the binary. As shown by the blue and green curves in Fig.~\ref{fig:k38f-planet}, 
at this state (the evolutionary times of 1700 and 1900 years), the semi-major axis of the planet remains constant and its 
eccentricity increases. Subsequently, the planet leaves the resonance migrating further inward until its reaches an orbit with a
semi-major axis of $a \approx 0.415$ AU between the 5:1 and 4:1 resonances where it then becomes stable.
At this state, the planet’s eccentricity reduces to an averaged value of $e \approx 0.03$. 
Compared to the semi-major axis and eccentricity of Kepler-38b (see Table~\ref{tab:kepler38}), 
the planet in the radiative disk model is closer to the binary, however, its eccentricity is in a similar range.
This mismatch between the observed value of the planet’s semi-major axis and that obtained from our simulations may be due to a 
too high a disk mass.

\begin{figure}
\center
\includegraphics[width=0.45\textwidth]{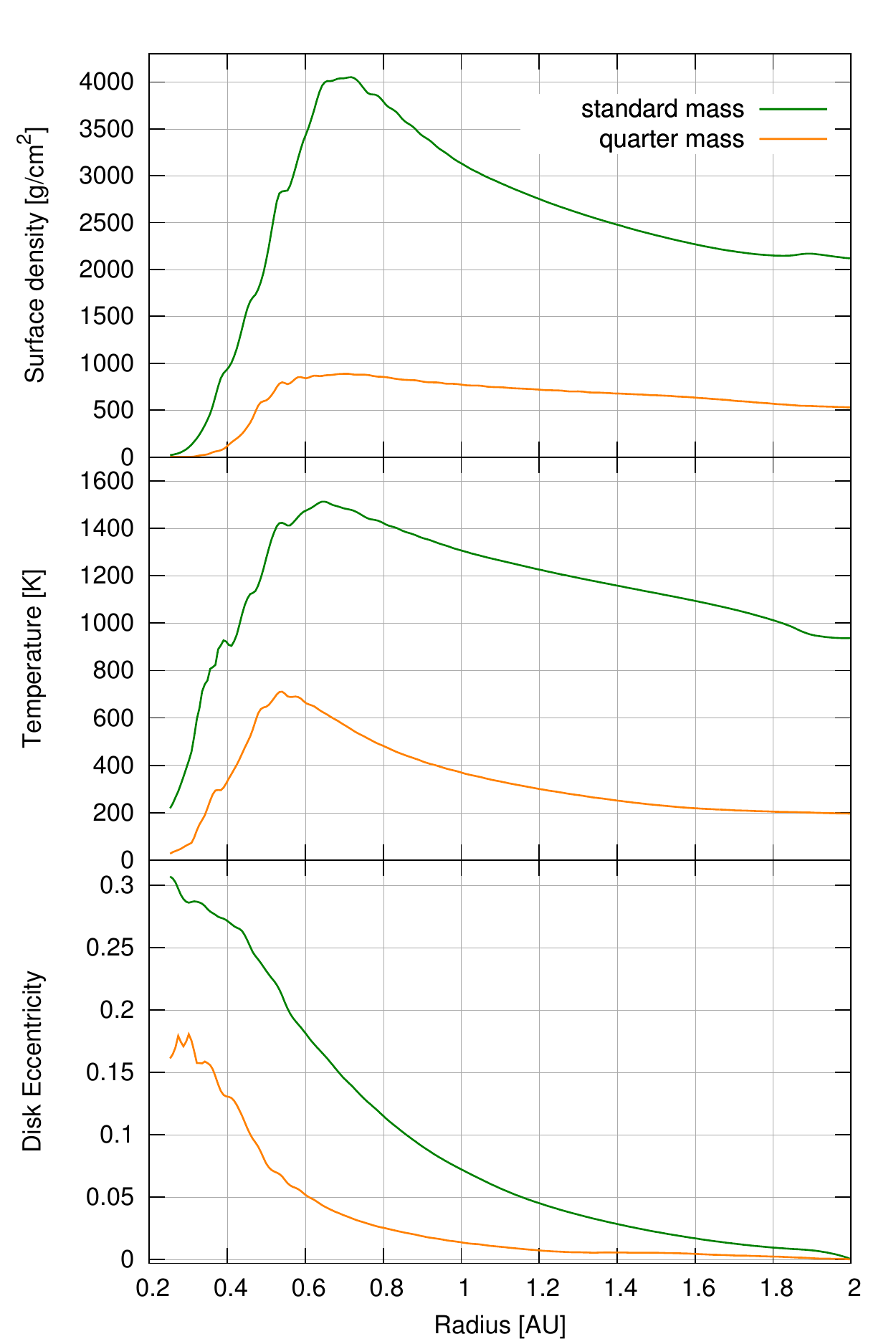} \\
\caption{Graphs of the surface density (in g/cm$^2$, top panel), temperature (in K, middle panel) and 
eccentricity (bottom panel) of two radiative disks with different masses. The green curve corresponds to
the same disk model as the green curve in Fig.~\ref{fig:k38f-structure} with the standard disk mass and 
viscosity ($\alpha = 0.01$). The orange curve shows a disk model with a mass equal to a quarter of the 
mass of our standard disk model and a lower disk viscosity of $\alpha = 0.004$.  
 }
\label{fig:k38f7v-structure}
\end{figure}

\begin{figure}
\center
\includegraphics[width=0.45\textwidth]{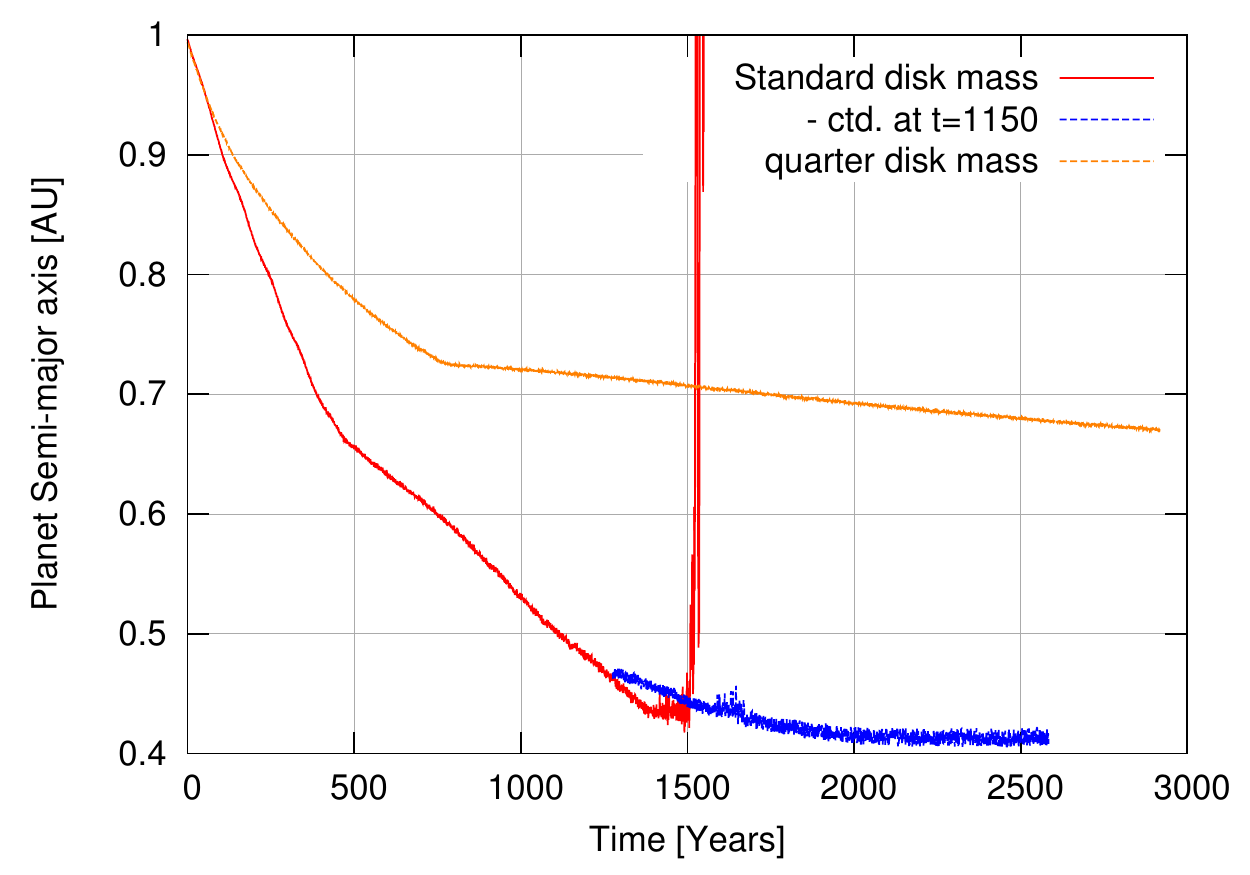} \\
\caption{The evolution of the semi-major axis of an embedded planet in a radiative disk with different masses.
The red and blue curves correspond to the standard disk mass model (as shown in Fig.~\ref{fig:k38f-planet}) and
the orange curve corresponds to a model with a quarter disk mass.
 }
\label{fig:k38f7v4-planet}
\end{figure}

\subsection{Planets in radiative disks with reduced mass and lower viscosity}
\label{subsec:rad_lowmass}
As indicated by the results of our simulations, the final location of a planet in a circumbinary disk depends on the 
location of the disk’s inner edge. In the radiative model with the standard disk mass, the inner edge of the disk
was quite close to the binary star. As a result, the planet was able to cross the 5:1 resonance, or became
unstable after being temporarily captured in that resonance. A survey of the currently known circumbinary planets 
indicates that in contrast to these results, all the observed planets are in orbits 
outside the 5:1 resonance. To ensure that the final location of the planet in our simulations will be farther away,
we need to develop a disk model with a large inner cavity such that the inner edge of the disk will be at a large
distance from the binary star. Two effects will be able to accommodate such a disk model; a reduced temperature, 
and a lower viscosity. 

In a radiative model, the midplane temperature of the disk is determined by the disk opacity, which
can be easily adjusted by the mass-content of the disk.
To simulate the evolution of the disk and its embedded planet for lower values of the disk temperature,
we, therefore, consider a model with a smaller disk mass. Additionally, we reduce the value of the disk viscosity. 
Specifically, we consider a disk with a quarter of the mass (i.e. a model in which the surface density at the outer boundary 
was set to a quarter of the original value), and
a smaller viscosity parameter, $\alpha = 0.004$. This value of $\alpha$ may be more realistic for protoplanetary disks 
in their final phases. In Fig.~\ref{fig:k38f7v-structure}, we compare the structure of this disk with 
that of our standard radiative  disk model. From top to bottom, the panels shows the radial dependence of the disk surface density, 
its temperature, and the disk eccentricity at the equilibrium state. The green curves correspond to the quantities in Fig.~\ref{fig:k38f-structure}.
Because of the disk’s lower mass, the surface density of the disk is naturally smaller. This reduces the disk’s optical depth and leads 
to a lower temperature, which subsequently results in a smaller disk thickness $H$. In agreement with previous results, a smaller $H$
results in a smaller eccentricity for the disk.
 
\begin{figure}
\center
\includegraphics[width=0.45\textwidth]{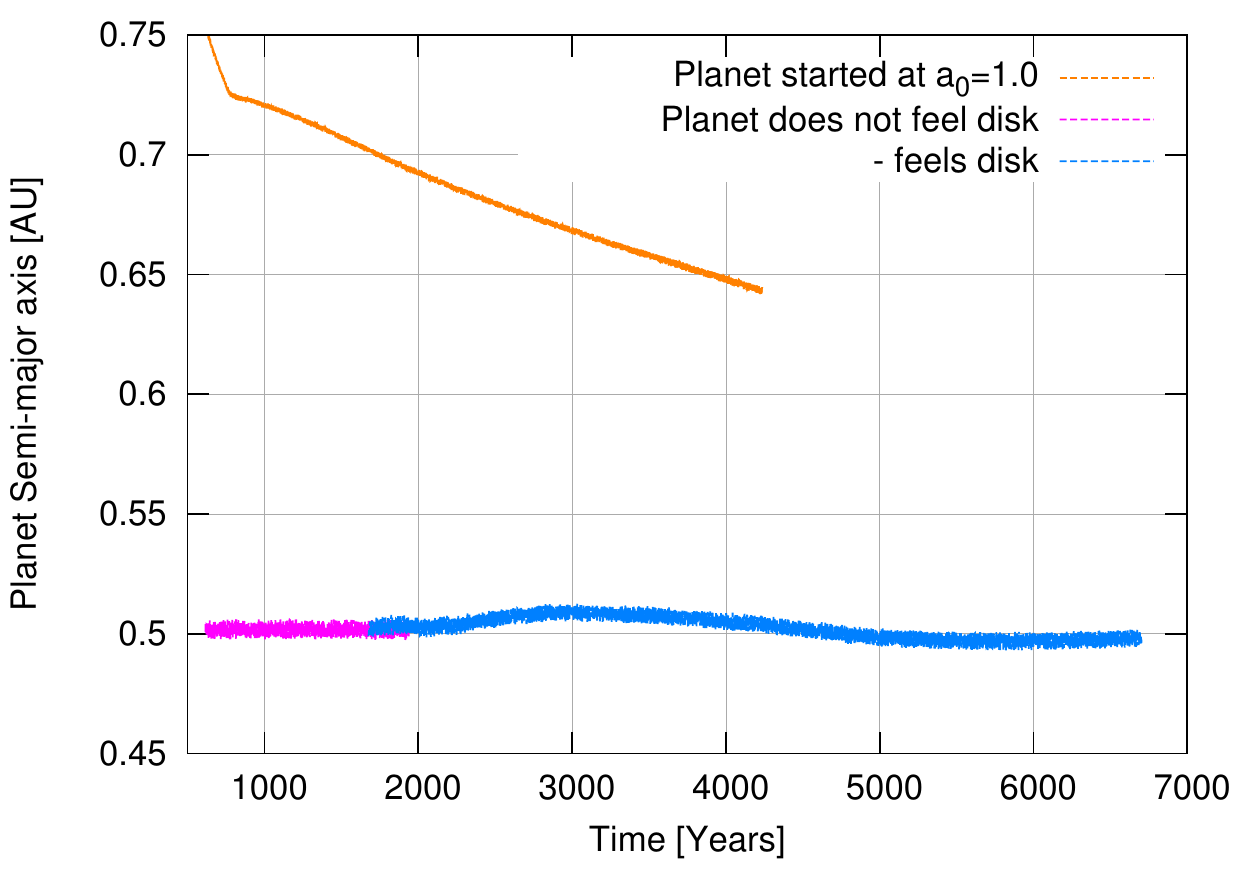} \\
\includegraphics[width=0.45\textwidth]{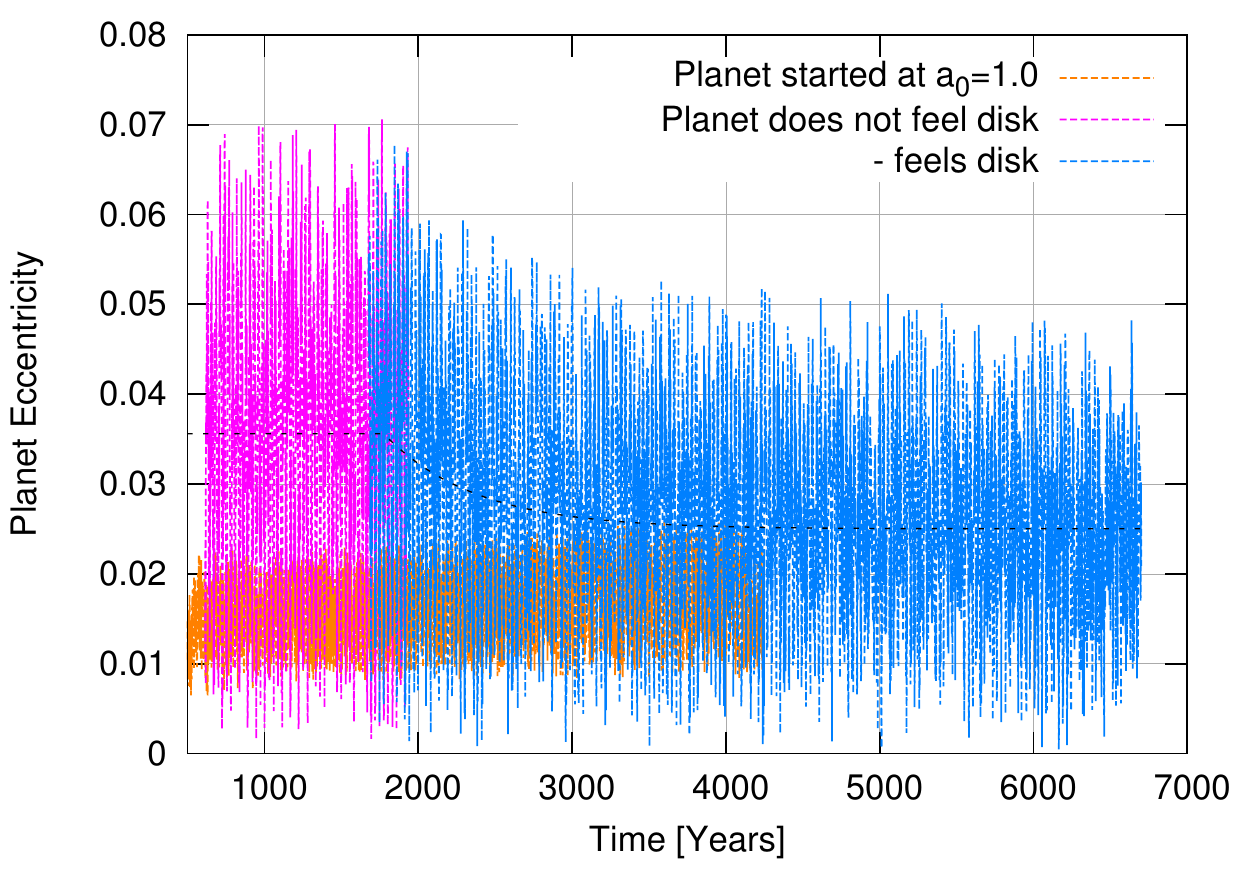}
\caption{the evolution of the semi-major axis and eccentricity of a planet embedded in a radiative disk
with a mass lower than that of our standard disk model. The orange curve corresponds to the same model as the orange curve in
Fig.~\ref{fig:k38f7v4-planet} with the planet started at $a_0 = 1.0$AU. The purple curve corresponds to a system in which
the planet was embedded in the disk at $a_0=0.5$ AU and it could move only under the influence of the gravity of the binary.
The light blue curve shows the same planet as in the purple curve. Except that in this case, at time $t=1700$, the effect of 
the disk was also taken into account. The dashed line in the bottom panel shows the evolution of the planet’s mean eccentricity. 
As shown by this panel, the first equilibrium is reached at $e \approx 0.036$.
Under the action of the disk, this value of the eccentricity exponentially decreases to a new mean value of $e \approx 0.026$  
in a timespan of $600$ years.
 }
\label{fig:k38f7v7-planet}
\end{figure}

We followed our standard procedure, i.e. after allowing the new disk (with lower mass and viscosity) to reach 
equilibrium, we placed a planet in a circular orbit at $a_0 = 1$AU. Fig.~\ref{fig:k38f7v4-planet} shows the 
evolution of the semi-major axis of the planet for the new quarter-mass model (orange line) in comparison to the previous 
simulations. As shown here, in both models, the planet migrates inward and the rate of migration 
is initially similar in both cases. However, despite the fact that
the reduced density would slow down the migration and the lower temperature would speed it up,
in the long run, the planet drifted inward at a much smaller rate than in the previous simulation.
In this reduced-mass disk model with the lower temperature and viscosity, the planet carves a deeper gap
where the central density in the vicinity of the planet is only about a quarter to a third of the ambient density
compared to the two-thirds in the previous model (see Fig.~\ref{fig:k38a5a-sig}).
This deeper gap brings the planet closer to the type-II migration regime,
and eventually leads to the very slow migration rate observed in Fig.~\ref{fig:k38f7v4-planet}, in the long run.
In this simulation, the migration is so slow that we could not follow the model to the final equilibrium state.

To overcome the issue with the slow migration, we ran an additional model where we placed the planet initially in an orbit with 
a smaller semi-major axis of $a_0=0.5$ AU. In contrast to the previous models, here we first 
evolve the system without the gravitational back-reaction of the disk, i.e.
the planet only feels the gravitational force of the two stars. In this way, we calculate 
the approximate equilibrium state of the combined disk+planet system, to avoid initial transients.
Once this equilibrium is reached, we include the gravitational back-reaction of the disk 
to allow for possible orbital evolution of the planet.
Fig.~\ref{fig:k38f7v7-planet} shows the results. 
The evolution of the planet semi-major axis is shown in the top panel and the bottom panel shows it eccentricity. 
The orange curve refers to the previous model where the planet was started
at $a_0=1.0$AU under the full action of the disk. As shown in the figure, while the planet semi-major axis is continuously 
shrinking, its eccentricity is slowly increasing. For the model started at $a_0 = 0.5$, the planet remains first in its orbit
while it does not feel the gravity of the disk (purple curve). At this state, the eccentricity of the planet oscillates 
between 0.0 and 0.07 with an average of about 0.036. We note that these variations are solely due to the effect of the binary star
and are always present for an orbiting planet.
Once the back-reaction of the disk is included, the orbital elements of the planet begin to slowly vary (light blue curve).
The planet begins to move slightly outward while its eccentricity becomes smaller due to the damping effect of the disk.
For illustration purposes, we have superimposed the black dashed line which in the first phase shows the equilibrium eccentricity, and
after the release of the planet, its exponential decline with a timescale of $600$yrs.
After its initial outward migration, the planet moves slowly inward and finally reaches a stable orbit exterior to the 5:1 resonance.
From the results shown in Fig.~\ref{fig:k38f7v7-planet}, 
we infer that the final position of the planet is very close to $a_\text{p} = 0.5$AU, just slightly outside its observed orbit.

\section{Summary and discussion}

Using 2D hydrodynamical simulations, we studied the evolution of a planet embedded in a circumbinary disk.
We carried out our simulations for the planetary system of Kepler-38 as our test binary. 
In our analysis, we adopted a two-step approach in which we first studied the structure of a circumbinary disk without 
a planet, and then included a planet in the disk and followed its evolution.
To connect to previous studies by other authors, we began by considering a locally isothermal equation of state in our disk
models. However, in order to extended the analysis to more realistic systems, we considered radiative models which
include viscous dissipation, vertical cooling, and radiative diffusion. In the following, we briefly present 
our most important results:\\

\noindent
a) Inner boundary \\
To enable mass accretion into the inner cavity of the disk, we used an open boundary condition for the disk inner edge.
This boundary condition allowed for free in-flow of material onto the binary star. At the outer edge of the disk, we 
considered a fixed-density boundary condition which, combined with the open boundary at the inner edge, resulted in disk 
models with constant mass-accretion rates. 
Compared to the disks with closed inner boundaries or those with slow viscous in-flows,
as used for example by \citet{2013A&A...556A.134P}, our open boundary condition 
lead to the disk models with reduced and broader maximum densities and smaller disk eccentricities. 
Initially, we considered the location of the inner edge of the computational grid , (i.e., $r_{min}$), to be at a distance
of approximately 1.7 times the binary semi-major axis. Following \citet{2013A&A...556A.134P},
who showed that the location of the disk inner edge can influence the structure of the inner part of the disk,
we carried out simulations for different values of $r_{min}$ and determined that a value of $r_{min} \approx  1.5 a_\text{bin}$ 
(equal to 0.22 AU for the Kepler-38 system) would be sufficient to capture all necessary physics.
A disk model with $r_{min}=0.20$ AU also resulted in the same gap profile.
We note that from a numerical standpoint, the largest possible value of $r_{min}$ should be considered. That is
because smaller values of this quantity will lead to significantly reduced timesteps in which case the FARGO-algorithm is no 
longer applicable. Therefore, in systems where the (expected) inner cavity is larger, a larger value of $r_{min}$ is certainly preferable.\\ 

\noindent
b) Planets in isothermal disks \\
We studied locally isothermal disks with a viscosity parameter $\alpha=0.01$ and vertical scale height of $H/r = 0.05$.
Before embedding the planet in the disk, we allowed the disk to reach a quasi-equilibrium state, which for our standard
disk model took approximately 2000 yrs ($\sim$39,000 binary orbits). 
We then started the planet at various distances from the binary and showed that as expected for a dissipative viscous disk,
the planet stops its radial migration and settles in the same orbit near the inner edge of the disk regardless of its starting point. 
For the case with $r_{min}=0.25$ AU, we found that the planet was captured in a 5:1 resonance with the binary and remained
in that orbit until the end of the simulation. The stability of the resonant configuration may be caused by the damping 
action of the disk, because circumbinary resonances such 5:1 are known to be unstable for planetary orbits \citep{1986A&A...167..379D}.
Hence, we expect unstable behaviour in this case once the disk has been dissipated,
possibly leading to encounters with the binary and departure from a closed orbit.
For a disk with $r_{min}=0.22$ AU (a more realistic case), the inner cavity was larger and as a result, the planet ceased its migration
slightly outside the 5:1 resonance at $a \approx 0.47$ AU, which is in a very good agreement with the observed location of circumbinary
planet Kepler-38b. The mean eccentricity of this planet, $e\approx 0.03$, also agrees with that of Kepler-38b.
Our results indicated that in the case of isothermal disks, the disk mass does not influence the final position of the planet
because the zero-torque position only depends on the density profile and not the absolute value, as the density scales out of the 
equations.\\

\noindent
c) Planets in radiative disks \\
In a final set of simulations, we developed more realistic disk models by including viscous dissipation, radiative
cooling, and diffusion. We found that when we used the same absolute density as in the standard isothermal disk ($\approx$ 2100 g/cm$^2$ at 2AU),
the viscous heating made the disk hotter compared to the isothermal cases. This additional heating caused the disk to
become more eccentric and the inner cavity to become smaller. As a result, the embedded planet was able to cross the 
5:1 resonance and reach its final position at $a\approx 0.415$ AU (slightly outside the instability region)
with a small eccentricity of $e\approx 0.03$. Such an orbit will again be unstable upon dissipation of the disk.

A reduction of the disk density combined with a lower viscosity lead to a wider inner cavity with a smaller disk temperature.
In this case, the disk mass becomes very small and as a result, the migration time becomes too long to 
follow the planet’s evolution to the end, as found by \citet{2013A&A...556A.134P} as well.
To save computational effort, we started a model with a planet initially much further in, 
and obtained an equilibrium state where the planet stopped its radial migration
just inside of $a \approx 0.5$ AU, again very close to the observed location.\\

In agreement with previous simulations \citep{2007A&A...472..993P,2013A&A...556A.134P}, our study indicates that  
planets migrate inward in circumbinary disks. The final position of a planet depends on the size of the inner cavity.
The results of isothermal runs suggest that the disk height has to be around $H/r = 0.05$, however,
no other information on the disk mass can be inferred from these simulations.
In fully radiative simulations, the mass of the disk will determine the temperature and hence $H/r$, as in these
systems the cooling depends on the vertical column density.
From our simulations we infer that the required surface density is probably around a quarter of the minimum solar mass
nebula.

To improve the models, the stellar irradiation from the two stars of the binary can be included.
The combined irradiation will heat up the inner parts of the disk causing the disk’s thickness to increase 
which itself results in an increase in the disk's eccentricity. What effects this may have on the final outcome 
of the simulations remains to be seen.
A further improvement can be obtained with full three-dimensional simulations including radiative transport.
However, considering the necessary long evolution time, where ten-thousands of binary orbits have to be followed,
3D simulations are certainly still challenging.

\begin{acknowledgements}
We thank Arnaud Pierens for stimulating discussions.
N.H. acknowledges support from the NASA ADAP grant NNX13AF20G, and the Alexander von Humboldt Foundation.
N.H. Would also like to thank the Computational Physics Group the University of T\"ubingen for their kind
hospitality during the course of this project.

\end{acknowledgements}
\bibliography{references}{}
\bibliographystyle{aa}
\end{document}